\documentclass[preprint,12pt]{elsarticle}

\usepackage{graphicx}
\usepackage[utf8]{inputenc}
\usepackage{amssymb}
\usepackage{amsmath}
\usepackage{amsfonts}
\usepackage{natbib}

\usepackage[colorlinks=true,linkcolor=blue,urlcolor=blue]{hyperref}
\usepackage{pygmentizedefs0}

\def\figdir{.}

\newcommand{\smatrix}[2]{\left#2\begin{array}{#1}}
\newcommand{\ematrix}[1]{\end{array}\right#1}
\def\pdiff#1#2{\frac{\partial {#1}}{\partial {#2}}}
\def\Hcurl{\vec{H}(\mathrm{curl})}
\def\Hdiv{\vec{H}(\mathrm{div})}

\def\Ab{\vec{A}}
\def\Pb{\vec{P}}
\def\eb{\vec{e}}
\def\db{\vec{d}}
\def\fb{\vec{f}}
\def\gb{\vec{g}}
\def\hb{\vec{h}}
\def\nb{\vec{n}}
\def\ub{\vec{u}}
\def\vb{\vec{v}}
\def\veps{\varepsilon}
\def\vphi{\varphi}
\def\vrho{\varrho}
\def\sigmab{\boldsymbol\sigma}
\def\omegab{\boldsymbol\omega}
\def\Pib{\boldsymbol\Pi}
\def\dV{\mbox{d}V}
\def\dS{\mbox{d}S}
\def\eeb#1{\eb\left(#1\right)}
\def\Hspace{\vec{H}_\#^1}
\def\Uspace{\boldsymbol{\mathcal U}}

\def\SfePy{\textit{SfePy}}

\begin{document}

\begin{frontmatter}

\title{Multiscale finite element calculations in Python using SfePy}

% \author{Robert Cimrman \and Vladimír Lukeš \and Eduard Rohan}

% \institute{%
% R.~Cimrman (ORCID:0000-0002-2660-545X)\at New Technologies - Research Centre, University of West
% Bohemia, Univerzitn\'{\i}~8, 30614, Pilsen, Czech Republic,
% \email{cimrman3@ntc.zcu.cz}
% \and
% V.~Luke\v{s} (ORCID:0000-0001-6579-1868) \at NTIS -- New Technologies for the Information
% Society, Faculty of Applied Sciences, University of West Bohemia,
% Univerzitn\'{\i}~8, 30614, Pilsen, Czech Republic, \email{vlukes@ntis.zcu.cz}
% \and
% E.~Rohan (ORCID:0000-0002-0300-4600) \at NTIS -- New Technologies for the Information
% Society, Faculty of Applied Sciences, University of West Bohemia,
% Univerzitn\'{\i}~8, 30614, Pilsen, Czech Republic, \email{rohan@ntis.zcu.cz}
% }

\author[NTC]{R.~ Cimrman}
\ead{cimrman3@ntc.zcu.cz}

\author[NTIS]{V.~Luke\v{s}}
\ead{vlukes@ntis.zcu.cz}

\author[NTIS]{E.~Rohan}
\ead{vlukes@ntis.zcu.cz}

\address[NTC]{New Technologies - Research Centre, University of West
Bohemia, Univerzitn\'{\i}~8, 301~00, Pilsen, Czech Republic}

\address[NTIS]{NTIS -- New Technologies for the Information
Society, Faculty of Applied Sciences, University of West Bohemia,
Univerzitn\'{\i}~8, 301~00, Pilsen, Czech Republic}

\begin{abstract}
  SfePy (Simple finite elements in Python) is a software for solving various
  kinds of problems described by partial differential equations in one, two or
  three spatial dimensions by the finite element method. Its source code is
  mostly (85\%) Python and relies on fast vectorized operations provided by the
  NumPy package. For a particular problem two interfaces can be used: a
  declarative application programming interface (API), where problem
  description/definition files (Python modules)
  are used to define a calculation, and an imperative API, that can be used for
  interactive commands, or in scripts and libraries. After outlining the SfePy
  package development, the paper introduces its implementation, structure and
  general features. The components for defining a partial differential equation
  are described using an example of a simple heat conduction problem.
  Specifically, the declarative API of SfePy is presented in the example. To
  illustrate one of SfePy's main assets, the framework for implementing complex
  multiscale models based on the theory of homogenization, an example of a
  two-scale piezoelastic model is presented, showing both the mathematical
  description of the problem and the corresponding code.
%   \subclass{35Qxx \and 65N30 \and 65M60 \and 65Y05 \and 74S05}
\end{abstract}


\end{frontmatter}

\section{Introduction}
\label{sec:intro}

\SfePy{} (\url{http://sfepy.org}) is a software for solving systems of coupled
partial differential equations (PDEs) by the finite element method (FEM) in 1D,
2D and 3D. It can be viewed both as a black-box PDE solver, and as a Python
package which can be used for building custom applications. It is a
multi-platform (Linux, Mac OS X, Windows) software released under the New BSD
license --- the source code hosting, issue tracker and continuous integration
tools are available thanks to the GitHub development platform.

\SfePy{} has been employed by our group for a range of topics in multiscale
modelling in biomechanics and materials science, including multiscale models of
biological tissues (bone, muscle tissue with blood perfusion)
\cite{cimrman07:_model,rohan09:_multis,cimrman10:_two,rohan12:_multis_fe,rohan12:_multis,rohan12:_hierar},
a fish heart model with active contraction \cite{Kochova_2015}, computations of
acoustic transmission coefficients across interfaces of arbitrary
microstructure geometry \cite{rohan10:_homog}, computations of phononic band
gaps \cite{rohan09:_numer,rohan11:_band_reiss_mindl}, the finite element
formulation of the Schroedinger equation
\cite{ptcp,IGA-FEM-Cimrman-1p,IGA-FEM-Cimrman-2}, and other applications.

In Section~\ref{sec:development} the programming language choice is discussed
and the \SfePy{} project development is described. In
Section~\ref{sec:description} an overview of the package is given, and a simple
example definition (a heat conduction problem) is presented.
Section~\ref{sec:hom_engine} introduces the \SfePy{}'s homogenization engine
--- a sub-package for defining complex multiscale problems using a simple
domain specific language, based on Python's dictionaries. Finally, in
Section~\ref{sec:piezo_example} a complex example --- a multiscale numerical
simulation of a piezoelectric structure --- is shown as expressed in the
homogenization engine syntax.

\section{Development}
\label{sec:development}

The code has been written primarily in the Python programming
language\footnote{\SfePy{} sources (version 2018.3) GitHub statistics: Python
  85.1\%, C 14.6\%, other: 0.3 \%.}. Python is a very high-level interpreted
programming language, that has a number of features appealing to scientists
(non-IT), such as: a clean, easy-to-read syntax, no manual memory management, a
huge standard library, a very good interoperability with other languages (C,
fortran), and a large and friendly scientific computing community. It allows
both fast exploration of various ideas and efficient implementation, thanks to
many high-performance solvers with a Python interface, and numerical tools and
libraries available among open source packages.

There are many finite element packages, commercial or open source, that can be
used from Python, and some of them use Python as a primary
language, notably Fenics \cite{fenics} or Firedrake \cite{firedrake}. This
indicates viability of our choice and is in agreement with our positive
experience with the language.

The \SfePy{} project uses Git \cite{git} for source code management and GitHub
development platform \cite{github} for the source code hosting and developer
interaction, see \url{https://github.com/sfepy/sfepy}, similarly to many other
scientific Python tools. Travis CI \cite{travis-ci} is used (via GitHub) for
running automatic tests after every uploaded commit. The developers and users
of the software can communicate using the mailing list
``sfepy@python.org''.
%\footnote{Kindly hosted at \url{https://mail.python.org/mm3/mailman3/lists/sfepy.python.org/}.}
The source code and the package usage and development are documented with the
help of the Sphinx documentation generator \cite{sphinx} that is also used to
generate the pages of the \SfePy{} project web-site.

The version 2018.3 has been released in September 17, 2018, and its
git-controlled sources contained 897 total files, 721447 total lines (1539277
added, 817830 removed) and 6386 commits done by 24 authors\footnote{Most of the
  authors contributed only one or a few commits.}. About 120000 lines (16 \%)
are the source code, the other lines belong to the finite element meshes,
documentation etc.

\section{Description}
\label{sec:description}

In this section we briefly outline the package implementation, structure and
general features. The components for defining a PDE to solve are described
using a simple example in Section~\ref{sec:description_example}.

\subsection{Performance due to scientific Python ecosystem}

Because Python is an interpreted language, and the standard implementation
(CPython) has slow loops, several approaches are used in the code to achieve
good (C-like) performance. For speed in general, it relies on fast vectorized
operations provided by NumPy arrays \cite{oliphant07:_python_scien_comput},
with significant use of advanced features such as broadcasting. C and Cython
\cite{bradshaw:_cython} are used in places where vectorization is not possible,
or is too difficult/unreadable.

\SfePy{} relies on a number of packages of the scientific Python software
stack, namely: SciPy \cite{jones--:_scipy} for sparse matrices, solvers and
algorithms, Matplotlib \cite{hunter07:_matpl} for 2D plots, Mayavi
\cite{ramachandran11:_mayav} for 3D plots and simple post-processing GUI,
PyTables \cite{pytables} for HDF5 file format support, SymPy \cite{sympy} for
symbolic operations/code generation, igakit (a part of PetIGA \cite{petiga})
for working with NURBS bases of the isogeometric analysis \cite{IGA-1,IGA-2}
etc.

Besides the vectorized operations of NumPy, other performance gains are enabled
by using very efficient solvers with a Python interface, such as UMFPACK
\cite{davis04:_algor} + scikit-umfpack \cite{scikit-umfpack}, MUMPS
\cite{mumps} or PETSc \cite{petsc-user-ref}. \SfePy{} can run in parallel,
using PETSc + petsc4py \cite{petsc4py} and mpi4py \cite{mpi4py} packages. We
employ the separation of concerns strategy with respect to parallelism: most of
the \SfePy{} code is serial, and there is a single dedicated module in
\SfePy{}, that, together with the flexible way of computing weak form integrals
on any subdomain (see below), allows parallel assembling of the discrete
systems and their subsequent solution.

% \paragraph{Design Overview}
\subsection{Design overview}

In \SfePy{}, the equations are not given in a fully symbolic way, as in, for
example, Fenics or Firedrake projects\footnote{Both use the Unified Form
Language from Fenics.}, but a simpler approach is used: the \SfePy{} package
comes with a database of predefined terms. A \emph{term} is the smallest
unit that can be used to build \emph{equations}. It corresponds to a weak
formulation integral over a (sub)domain and takes usually several arguments:
(optional) material parameters, a single virtual (or test) function variable
and zero or more state (or unknown) variables. The available terms are listed
at our web site \cite{sfepy-web}, currently there are 118 terms.
%(\url{http://sfepy.org/doc-devel/terms_overview.html}).

The high-level code that handles a PDE discretization in \SfePy{} is independent
of a method of domain or variable discretization. For each particular method of
discretization, there is a sub-package that implements the specific
functionality (degrees of freedom management, selection of subdomains,
reference domain mappings, etc.). This abstraction allows adding various
discretization methods. The following ones are currently implemented:
\begin{itemize}
\item the finite element method on 1D line, 2D area (triangle, rectangle) and
  3D volume (tetrahedron, hexahedron) finite elements; with two kinds of
  polynomial bases:
  \begin{itemize}
  \item the classical nodal (Lagrange) basis that can be used with all
    supported element/cell types;
  \item the hierarchical (Lobatto) basis
    \cite{solin03:_higher_order_finit_elemen_method} that can be used with
    tensor-product elements (line, rectangle, hexahedron).
  \end{itemize}
  The basis function polynomials of an arbitrary order (theoretically, see
  limitations below) as well as the corresponding quadrature rules are
  supported. The Lagrange basis is implemented in C/Cython, while the Lobatto
  basis functions use a C code generated using SymPy.
\item the isogeometric analysis \cite{IGA-1} with a NURBS or B-spline basis,
  implemented using the Bézier extraction approach \cite{IGA-2}, currently
  limited to single NURBS patch domains, see \cite{Cimrman_IGA_2014}.
\end{itemize}
All the basis functions listed above support the $H^1$ function spaces only
($\Hcurl$, $\Hdiv$ spaces are not currently implemented). In addition to the
above elements, two structural elements are implemented (using \SfePy{} terms):
the hyperelastic Mooney-Rivlin membrane \cite{Wu_Du_Tan_1996}, and the shell10x
element term based on the Reissner-Mindlin theory
\cite{Zemcik_Rolfes_Rose_Tessmer_2006}.

\subsection{Working with \SfePy{}}

\SfePy{} can solve many problems described by PDEs in the weak form. For a
particular problem, there are two interfaces that can be used:
\begin{itemize}
\item a declarative API, where problem description/definition files (Python
  modules) are used to define a calculation;
\item an imperative API, that can be used for interactive commands, or in
  scripts and libraries.
\end{itemize}
Both the above APIs closely correspond to the mathematical description of the
weak form PDEs. An advanced use of the declarative API is demonstrated in the
piezoelectric model example in Section~\ref{sec:piezo_example}.

The declarative API involves almost no programming besides using basic Python
data types (dicts, lists, tuples, strings, etc.) and allows a lazy definition
of the problem, called \emph{problem configuration}, as well as a manipulation
with the problem configuration. Prior to a problem solution, the problem
configuration is automatically translated into a problem object using the
imperative API. The \SfePy{} package contains several top-level scripts that
can be used to run simulations defined using the declarative API. The two
common ones are:
\begin{itemize}
\item the \texttt{simple.py} script that allows running regular calculations of
  PDEs,
\item the \texttt{homogen.py} script that allows running the homogenization
  engine to compute effective material parameters, see
  Section~\ref{sec:hom_engine}.
\end{itemize}

The imperative API allows immediate evaluation of expressions, and thus
supports interactive exploration or inspection of the FE data. It is also more
powerful than the declarative API as a user is free to perform non-predefined
tasks. The problems defined using the imperative API usually have a
\texttt{main()} function and can be run directly using the Python interpreter.

In the both cases, a problem definition is a Python module, so all the power of
Python (and supporting \SfePy{} modules) is available when needed for complex
problems.

% \paragraph{Simple Example: Heat Conduction}
\subsection{Simple example: heat conduction}
\label{sec:description_example}

Systems of PDEs are defined using keywords or classes corresponding to
mathematical objects present in the weak formulation of the PDEs. Here we
illustrate the components of the problem definition using a simple example. We
wish to solve a heat conduction problem, that can be written in the weak form
as follows: Find the temperature $u \in H^1(\Omega)$ such that for all $v \in
H^1_0(\Omega)$ holds
\begin{displaymath}
  \int_{\Omega} v \pdiff{u}{t}
  + \int_{\Omega} c \nabla v \cdot \nabla u
  = 0
  \;, \forall v \;, u(x, 0) = g(x) \;, u(x, t) =
  \smatrix{rl}{\{}-2 & x \in \Gamma_{\mathrm{left}} \;, \\ 2 & x \in
  \Gamma_{\mathrm{right}} \;, \ematrix{.}
\end{displaymath}
where $c$ is a material parameter (a thermal diffusivity). Below we show the
declarative way of defining the ingredients necessary to solve this problem
with \SfePy{}. For complete examples illustrating both the declarative and
imperative APIs, see the accompanying dataset \cite{sfepy-examples-zenodo}.
\begin{itemize}
\item The domain $\Omega$ has to be discretized, resulting in a finite element
  \textbf{mesh}. The mesh can be loaded from a file (generated by external
  tools) or generated by the code (simple shapes).

  \begin{Verbatim}[commandchars=\\\{\}]
\PY{n}{filename\PYZus{}mesh} \PY{o}{=} \PY{l+s+s1}{\PYZsq{}}\PY{l+s+s1}{meshes/3d/cylinder.mesh}\PY{l+s+s1}{\PYZsq{}}
\end{Verbatim}

\item \textbf{Regions} serve as domains of integration and allow defining
  boundary and initial conditions. Subdomains of various topological dimension
  can be defined. The mesh/domain and region handling uses a C data structure
  adapted from \cite{Logg_2009}. The following code defines the domain $\Omega$
  and the boundaries $\Gamma_{\mathrm{left}}$, $\Gamma_{\mathrm{right}}$.

  \begin{Verbatim}[commandchars=\\\{\}]
\PY{n}{regions} \PY{o}{=} \PY{p}{\PYZob{}}
    \PY{l+s+s1}{\PYZsq{}}\PY{l+s+s1}{Omega}\PY{l+s+s1}{\PYZsq{}} \PY{p}{:} \PY{l+s+s1}{\PYZsq{}}\PY{l+s+s1}{all}\PY{l+s+s1}{\PYZsq{}}\PY{p}{,}
    \PY{l+s+s1}{\PYZsq{}}\PY{l+s+s1}{Left}\PY{l+s+s1}{\PYZsq{}} \PY{p}{:} \PY{p}{(}\PY{l+s+s1}{\PYZsq{}}\PY{l+s+s1}{vertices in (x \PYZlt{} 0.00001)}\PY{l+s+s1}{\PYZsq{}}\PY{p}{,} \PY{l+s+s1}{\PYZsq{}}\PY{l+s+s1}{facet}\PY{l+s+s1}{\PYZsq{}}\PY{p}{)}\PY{p}{,}
    \PY{l+s+s1}{\PYZsq{}}\PY{l+s+s1}{Right}\PY{l+s+s1}{\PYZsq{}} \PY{p}{:} \PY{p}{(}\PY{l+s+s1}{\PYZsq{}}\PY{l+s+s1}{vertices in (x \PYZgt{} 0.099999)}\PY{l+s+s1}{\PYZsq{}}\PY{p}{,} \PY{l+s+s1}{\PYZsq{}}\PY{l+s+s1}{facet}\PY{l+s+s1}{\PYZsq{}}\PY{p}{)}\PY{p}{,}
\PY{p}{\PYZcb{}}
\end{Verbatim}

\item \textbf{Fields} correspond to the discrete function spaces and are
  defined using the \textit{(numerical data type, number of components, region
    name, approximation order)} tuple. A field can be defined on the whole
  domain, on a volume (cell) subdomain or on a surface (facet) region.

  \begin{Verbatim}[commandchars=\\\{\}]
\PY{n}{fields} \PY{o}{=} \PY{p}{\PYZob{}}
    \PY{l+s+s1}{\PYZsq{}}\PY{l+s+s1}{temperature}\PY{l+s+s1}{\PYZsq{}} \PY{p}{:} \PY{p}{(}\PY{l+s+s1}{\PYZsq{}}\PY{l+s+s1}{real}\PY{l+s+s1}{\PYZsq{}}\PY{p}{,} \PY{l+m+mi}{1}\PY{p}{,} \PY{l+s+s1}{\PYZsq{}}\PY{l+s+s1}{Omega}\PY{l+s+s1}{\PYZsq{}}\PY{p}{,} \PY{l+m+mi}{1}\PY{p}{)}\PY{p}{,}
\PY{p}{\PYZcb{}}
\end{Verbatim}

\item The fields (FE spaces) can be used to define \textbf{variables}.
  Variables come in three flavors: \texttt{unknown field} for state variables,
  \texttt{test field} for test (virtual) variables and \texttt{parameter field}
  for variables with known values of degrees of freedom (DOFs). The definition
  items for an unknown variable definition are: \textit{(\texttt{'unknown
      field'}, field name, order in global vector, [optional history size])}.
  In the snippet below, the history size is 1, as the previous time step state
  is required for the numerical time derivative. For a test variable, the last
  item is the name of the corresponding unknown variable.

  \begin{Verbatim}[commandchars=\\\{\}]
\PY{n}{variables} \PY{o}{=} \PY{p}{\PYZob{}}
    \PY{l+s+s1}{\PYZsq{}}\PY{l+s+s1}{u}\PY{l+s+s1}{\PYZsq{}} \PY{p}{:} \PY{p}{(}\PY{l+s+s1}{\PYZsq{}}\PY{l+s+s1}{unknown field}\PY{l+s+s1}{\PYZsq{}}\PY{p}{,} \PY{l+s+s1}{\PYZsq{}}\PY{l+s+s1}{temperature}\PY{l+s+s1}{\PYZsq{}}\PY{p}{,} \PY{l+m+mi}{0}\PY{p}{,} \PY{l+m+mi}{1}\PY{p}{)}\PY{p}{,}
    \PY{l+s+s1}{\PYZsq{}}\PY{l+s+s1}{v}\PY{l+s+s1}{\PYZsq{}} \PY{p}{:} \PY{p}{(}\PY{l+s+s1}{\PYZsq{}}\PY{l+s+s1}{test field}\PY{l+s+s1}{\PYZsq{}}\PY{p}{,}    \PY{l+s+s1}{\PYZsq{}}\PY{l+s+s1}{temperature}\PY{l+s+s1}{\PYZsq{}}\PY{p}{,} \PY{l+s+s1}{\PYZsq{}}\PY{l+s+s1}{u}\PY{l+s+s1}{\PYZsq{}}\PY{p}{)}\PY{p}{,}
\PY{p}{\PYZcb{}}
\end{Verbatim}

\item \textbf{Materials} correspond to all parameters defined point-wise in
  quadrature points, that can be given either as constants, or as general
  functions of time and quadrature point coordinates. Here we just define the
  constant parameter $c$, as a part of the material \texttt{'m'}.

  \begin{Verbatim}[commandchars=\\\{\}]
\PY{n}{materials} \PY{o}{=} \PY{p}{\PYZob{}}
    \PY{l+s+s1}{\PYZsq{}}\PY{l+s+s1}{m}\PY{l+s+s1}{\PYZsq{}} \PY{p}{:} \PY{p}{(}\PY{p}{\PYZob{}}\PY{l+s+s1}{\PYZsq{}}\PY{l+s+s1}{c}\PY{l+s+s1}{\PYZsq{}} \PY{p}{:} \PY{l+m+mf}{1.0e\PYZhy{}5}\PY{p}{\PYZcb{}}\PY{p}{,}\PY{p}{)}\PY{p}{,}
\PY{p}{\PYZcb{}}
\end{Verbatim}

\item Similarly to materials, the Dirichlet (essential) \textbf{boundary
    conditions} can be defined using constants or general functions of time and
  coordinates. In our case we set the values of $u$ to 2 and -2 on
  $\Gamma_{\mathrm{left}}$ and $\Gamma_{\mathrm{right}}$, respectively.

  \begin{Verbatim}[commandchars=\\\{\}]
\PY{n}{ebcs} \PY{o}{=} \PY{p}{\PYZob{}}
    \PY{l+s+s1}{\PYZsq{}}\PY{l+s+s1}{u1}\PY{l+s+s1}{\PYZsq{}} \PY{p}{:} \PY{p}{(}\PY{l+s+s1}{\PYZsq{}}\PY{l+s+s1}{Left}\PY{l+s+s1}{\PYZsq{}}\PY{p}{,} \PY{p}{\PYZob{}}\PY{l+s+s1}{\PYZsq{}}\PY{l+s+s1}{u.0}\PY{l+s+s1}{\PYZsq{}} \PY{p}{:} \PY{l+m+mf}{2.0}\PY{p}{\PYZcb{}}\PY{p}{)}\PY{p}{,}
    \PY{l+s+s1}{\PYZsq{}}\PY{l+s+s1}{u2}\PY{l+s+s1}{\PYZsq{}} \PY{p}{:} \PY{p}{(}\PY{l+s+s1}{\PYZsq{}}\PY{l+s+s1}{Right}\PY{l+s+s1}{\PYZsq{}}\PY{p}{,} \PY{p}{\PYZob{}}\PY{l+s+s1}{\PYZsq{}}\PY{l+s+s1}{u.0}\PY{l+s+s1}{\PYZsq{}} \PY{p}{:} \PY{o}{\PYZhy{}}\PY{l+m+mf}{2.0}\PY{p}{\PYZcb{}}\PY{p}{)}\PY{p}{,}
\PY{p}{\PYZcb{}}
\end{Verbatim}

\item The \textbf{initial conditions} can be defined analogously, here we
  illustrate how to use a function. The conditions are applied in the whole
  domain $\Omega$. The code assumes NumPy was imported (\texttt{import numpy as
    np}), and \texttt{ic\_max} is a constant defined outside the function.

  \begin{Verbatim}[commandchars=\\\{\}]
\PY{k}{def} \PY{n+nf}{get\PYZus{}ic}\PY{p}{(}\PY{n}{coors}\PY{p}{,} \PY{n}{ic}\PY{p}{)}\PY{p}{:}
    \PY{n}{x}\PY{p}{,} \PY{n}{y}\PY{p}{,} \PY{n}{z} \PY{o}{=} \PY{n}{coors}\PY{o}{.}\PY{n}{T}
    \PY{k}{return} \PY{l+m+mi}{2} \PY{o}{\PYZhy{}} \PY{l+m+mf}{40.0} \PY{o}{*} \PY{n}{x} \PY{o}{+} \PY{n}{ic\PYZus{}max} \PY{o}{*} \PY{n}{np}\PY{o}{.}\PY{n}{sin}\PY{p}{(}\PY{l+m+mi}{4} \PY{o}{*} \PY{n}{np}\PY{o}{.}\PY{n}{pi} \PY{o}{*} \PY{n}{x} \PY{o}{/} \PY{l+m+mf}{0.1}\PY{p}{)}
\PY{n}{functions} \PY{o}{=} \PY{p}{\PYZob{}}
    \PY{l+s+s1}{\PYZsq{}}\PY{l+s+s1}{get\PYZus{}ic}\PY{l+s+s1}{\PYZsq{}} \PY{p}{:} \PY{p}{(}\PY{n}{get\PYZus{}ic}\PY{p}{,}\PY{p}{)}\PY{p}{,}
\PY{p}{\PYZcb{}}
\PY{n}{ics} \PY{o}{=} \PY{p}{\PYZob{}}
    \PY{l+s+s1}{\PYZsq{}}\PY{l+s+s1}{ic}\PY{l+s+s1}{\PYZsq{}} \PY{p}{:} \PY{p}{(}\PY{l+s+s1}{\PYZsq{}}\PY{l+s+s1}{Omega}\PY{l+s+s1}{\PYZsq{}}\PY{p}{,} \PY{p}{\PYZob{}}\PY{l+s+s1}{\PYZsq{}}\PY{l+s+s1}{u.0}\PY{l+s+s1}{\PYZsq{}} \PY{p}{:} \PY{l+s+s1}{\PYZsq{}}\PY{l+s+s1}{get\PYZus{}ic}\PY{l+s+s1}{\PYZsq{}}\PY{p}{\PYZcb{}}\PY{p}{)}\PY{p}{,}
\PY{p}{\PYZcb{}}
\end{Verbatim}

\item The PDEs can be built as a linear combination of many predefined terms.
  Each term has its quadrature order and its region of integration. The
  integral specifies a numerical quadrature order.

  \begin{Verbatim}[commandchars=\\\{\}]
\PY{n}{integrals} \PY{o}{=} \PY{p}{\PYZob{}}
    \PY{l+s+s1}{\PYZsq{}}\PY{l+s+s1}{i}\PY{l+s+s1}{\PYZsq{}} \PY{p}{:} \PY{l+m+mi}{2}\PY{p}{,}
\PY{p}{\PYZcb{}}
\PY{n}{equations} \PY{o}{=} \PY{p}{\PYZob{}}
    \PY{l+s+s1}{\PYZsq{}}\PY{l+s+s1}{Temperature}\PY{l+s+s1}{\PYZsq{}} \PY{p}{:} \PY{l+s+s2}{\PYZdq{}\PYZdq{}\PYZdq{}}\PY{l+s+s2}{dw\PYZus{}volume\PYZus{}dot.i.Omega(v, du/dt)}
\PY{l+s+s2}{                     + dw\PYZus{}laplace.i.Omega(m.c, v, u) = 0}\PY{l+s+s2}{\PYZdq{}\PYZdq{}\PYZdq{}}
\PY{p}{\PYZcb{}}
\end{Verbatim}

\item A complete example contains additional information (notably a
  configuration of solvers), see \cite{sfepy-examples-zenodo}. There, the above
  code snippets are used in the \texttt{heat\_cond\_declarative.py} file. The
  simulation is then launched using the \texttt{python <path/to/>simple.py
    heat\_cond\_declarative.py} command. In Fig.~\ref{fig:simple-results},
  illustrative results of the above computation are shown.
\end{itemize}
\begin{figure}[htp!]
  \centering
  \includegraphics[width=0.48\linewidth]{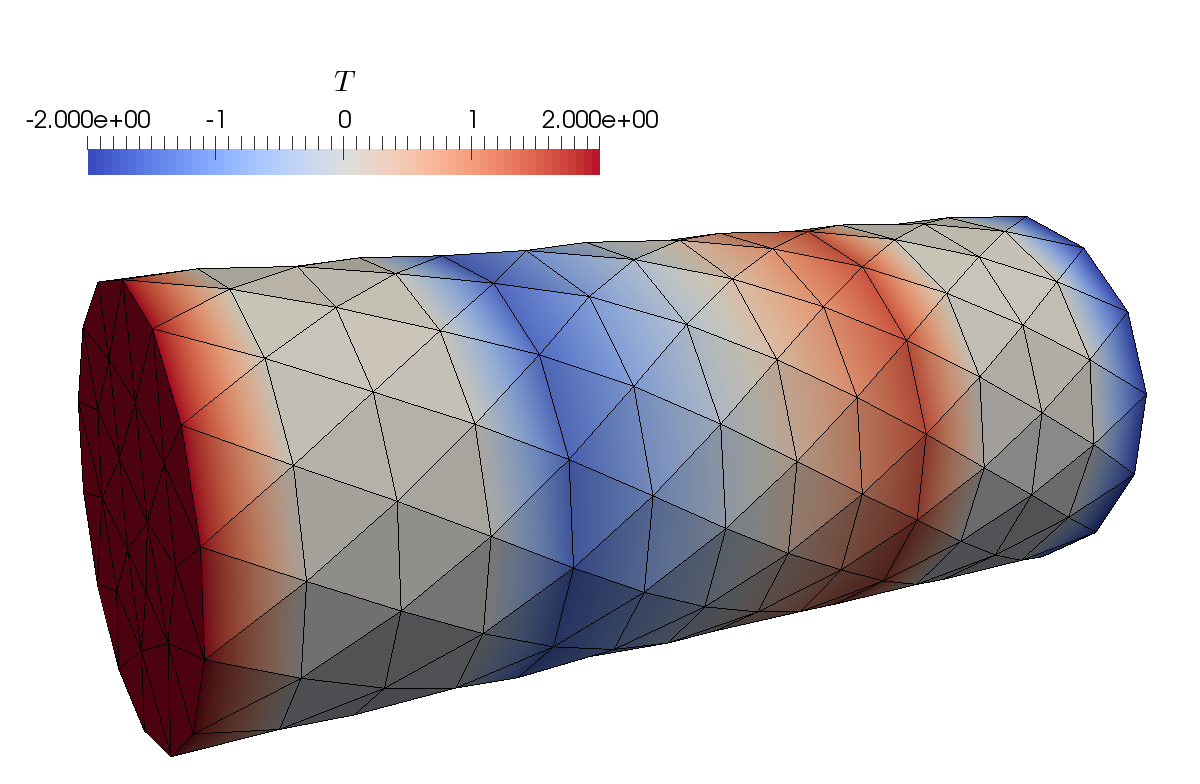}
  \hfill
  \includegraphics[width=0.48\linewidth]{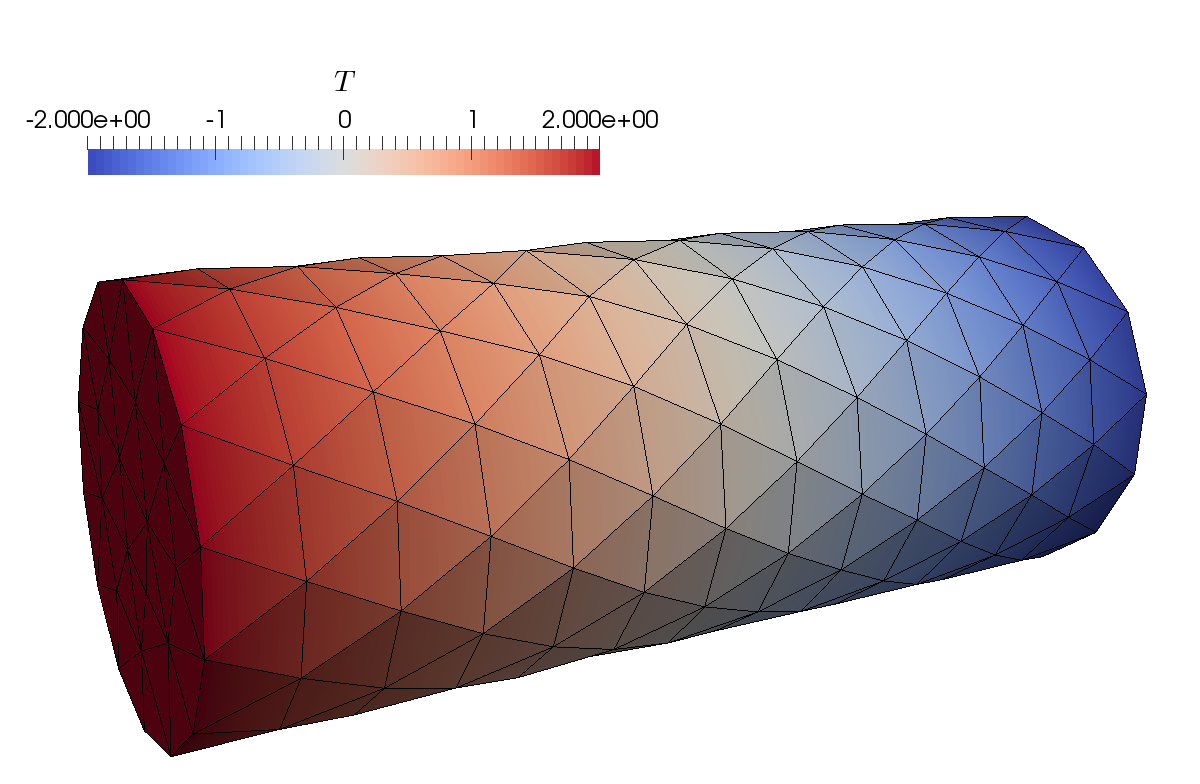}
  \caption{Initial and final snapshots of the temperature evolution.}
  \label{fig:simple-results}
\end{figure}

% More examples can be found at \SfePy{}'s web pages\cite{}.
%\url{http://docs.sfepy.org/gallery/gallery} or \url{http://sfepy.org/doc-devel/examples.html}.

\subsection{FE mesh handling and post-processing}
\label{sec:description_run}

\begin{figure}[htp!]
  \centering
  \includegraphics[width=0.85\linewidth]{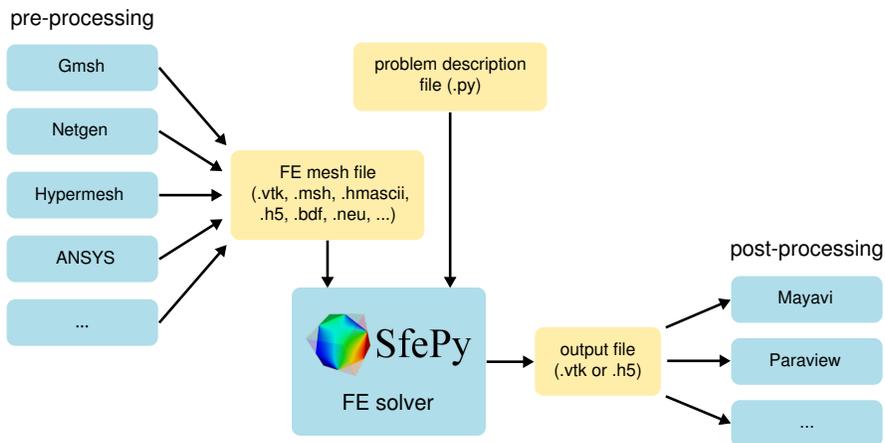}
  \caption{Data exchange between \SfePy{} and external pre-processing
  and post-processing tools.}
  \label{fig:data_workflow}
\end{figure}

\SfePy{} has no meshing capabilities besides several simple mesh generators (a
block mesh generator, an open/closed cylinder mesh generator, a mesh generator
from CT image data), but several operations like merging of matching meshes are
supported. The FE mesh needs to be provided in a file in one of the supported
formats, notably the legacy VTK format
\cite{kitware10:_visual_toolk_users_guide}. The results are stored in legacy
VTK files, or in, usually in case of time-dependent problems with many time
steps, custom HDF5 \cite{group10:_hierar} files. Many standard open-source
tools can be used to display the VTK files, namely Paraview
\cite{henderson07:_parav_guide_paral_visual_applic}, or Mayavi
\cite{ramachandran11:_mayav}, see data workflow of a simulation in
Fig.~\ref{fig:data_workflow}. Mayavi is supported directly within \SfePy{}
via the \texttt{postproc.py} script. The \texttt{extractor.py} script is
provided to extract/convert the HDF5 file content to VTK.

\subsection{Solvers}

\SfePy{} provides and uses a unified interface to many standard codes, for
example UMFPACK \cite{davis04:_algor}, MUMPS \cite{mumps}, PETSc
\cite{petsc-user-ref}, Pysparse \cite{geus:_pyspar_docum} as well as the
solvers available in SciPy. Various solver classes are supported:
time-stepping, nonlinear, linear, eigenvalue problem and optimization solvers.
An automatically generated list of all the supported solvers can be found at
the \SfePy{} web site \cite{sfepy-web}.
%\url{http://sfepy.org/doc-devel/users_guide.html#solvers}.

Besides external solvers, several solvers are implemented directly in \SfePy{},
for example:
\begin{itemize}
\item \texttt{ts.simple}: Implicit time stepping solver with a fixed time step,
  suitable also for quasistatic problems.
\item \texttt{ts.newmark}, \texttt{ts.bathe}, \texttt{ts.generalized\_alpha}:
  Solve elastodynamics problems by the Newmark, Bathe, generalized-$\alpha$
  methods, respectively.
\item \texttt{nls.newton}: The Newton nonlinear solver with a backtracking
  line-search.
\end{itemize}

A typical problem solution, when using the declarative problem definition API,
then involves calling a time-stepping solver that calls a nonlinear solver in
each time step, which, in turn, calls a linear solver in the nonlinear
iterations. A unified approach is used here: for stationary problems, a dummy
time-stepping solver (\texttt{ts.stationary}) is automatically created.
Similarly, a nonlinear solver is used to solve both the linear and non-linear
problems. This simplifies imposing non-homogeneous Dirichlet boundary
conditions: in the context of a nonlinear solver, the increment of the
constrained DOFs is always zero and the non-zero boundary values can be ignored
during the assembling.

\subsection{Limitations}

The limitations can be split into two groups. The first group is related to
limited number of developers and our research focus: certain features are
missing, because they do not fall into our field of research (e.g. the vector
finite elements). The limitations in the second group are more fundamental.
Because the code relies on vectorization provided by NumPy, the code tries to
work on all cells in a region in each operation: for example, all local finite
element matrices are evaluated in a vectorized way into a single large NumPy
array, and then assembled to a SciPy's sparse matrix. This places a restriction
on a practically usable order of the basis function polynomials, especially for
3D hexahedrons, where orders greater than 4 are not practically usable. Using
NumPy's arrays places another restriction: data homogeneity. So, for examples,
the FE basis polynomial order has to be uniform over the whole (sub)domain
where a field is defined. This is incompatible with an adaptive mesh
refinement, especially the $hp$-adaptivity. Note that \SfePy{} supports meshes
with level-1 hanging nodes, so a limited $h$-adaptivity is partly possible.

\section{Homogenization engine}\label{sec:hom_engine}

In this section we briefly outline the approach to solving multiscale problems
based on the theory of homogenization \cite{Allaire1992,Cioranescu1999book}.
The main asset of the homogenization approach is that a homogenized model can
take into account various details at the microstructure scale (topology,
heterogeneous material parameters, etc.) without actually meshing those
detailed features on a macroscopic domain, which would lead to an extremely
large problem.

The homogenization engine is a feature that is in our opinion unique to \SfePy{}.
It has been developed to allow an easy and flexible formulation of problems
arising from use of the homogenization theory applied to strongly
heterogeneous multiscale material models. Such models, as can be seen in the
non-trivial example in Section~\ref{sec:piezo_example}, can have a complicated data
flow and dependencies of various subproblems involved in the definition.

A typical homogenization procedure for a linear problem\footnote{The situation
  is much more complicated for nonlinear problems: a microproblem needs to be
  typically solved in every macroscopic integration point. This mode is also
  supported in \SfePy{}.} involves the
following steps:
\begin{enumerate}
\item Compute characteristic (corrector) functions by solving auxiliary
  corrector problems on a reference periodic cell domain that describes the
  microstructure (exactly or in a statistical sense).
\item Using the corrector functions, evaluate the homogenized
  coefficients. Those coefficients correspond to effective macroscopic
  properties of the material with the given microstructure as the
  microstructure characteristic scale tends to zero.
\item Solve the homogenized model with the obtained effective homogenized
  coefficients on a macroscopic domain.
\end{enumerate}
There can be a number of the auxiliary corrector problems as well as the
homogenized coefficients. The homogenization engine allows to describe their
relationships and the dependencies among them and resolves the problems in a
correct order automatically. A complete problem is described in one or more
problem definition files using the declarative API, and data dependencies are
described using Python dictionaries as a small domain-specific language.

The homogenization engine allows to solve microscopic subproblems and
evaluate homogenized coefficients in parallel with use of either multithreading
or multiprocessing features of a computer system. The distribution of
microproblems between multiple threads or CPUs is
governed by a function that puts the microproblems with resolved dependencies
into the work queue, collects solved microscopic solutions and updates
a dependency table according to the obtained results. Workers, i.e.
threads or CPU cores, solve tasks from the work queue until it is empty.
If the same microproblem needs to be solved multiple times with different
parameters, typically for nonlinear problems, the total amount of microscopic
tasks is divided into several chunks that are distributed to multiple
workers. In the case of multiprocessing, the MPI library is used to communicate
between computational nodes.

The parallel computation is crucial for nonlinear problems where microproblems
have to be resolved in all macroscopic integration points in many time or
iteration steps.

\section{Multiscale numerical simulation of piezoelectric structure}
\label{sec:piezo_example}

The complex multiscale model, solved by the means of the homogenization method,
and its implementation in \SfePy{} are presented in this section.

\subsection{Mathematical model of piezoelectric media}
\label{sec:piezo}

\begin{figure}
    \centering
    \includegraphics[width=0.7\linewidth]{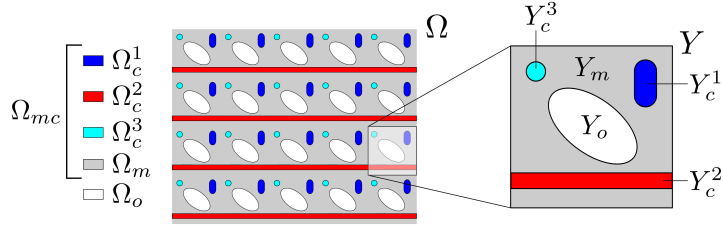}
    \caption{The scheme of the representative periodic cell decomposition and the
    generated periodic structure.}\label{fig-mac_mic}
\end{figure}

We consider a porous piezoelectric medium which consists of a
piezoelectric matrix, embedded metallic electrodes (conductors) and void
inclusions. These components are arranged in a periodic lattice so that the
medium can be generated by copies of the reference unit cell,
see~Fig.\ref{fig-mac_mic}. The mechanical behavior of such a structure can be
described using the two-scale asymptotic homogenization method, see
\cite{Allaire1992,Cioranescu1999book}. The quantities oscillating within the heterogeneous
structure with the period equal to the size of the periodic unit are labelled
by the superscript $^\veps$ in the subsequent text.

The mechanical properties of the piezoelectric solid are given by the following
constitutive equations
\begin{equation}\label{eq-constitutive}
    \begin{split}
        \sigma^\veps_{ij}(\ub^\veps,\vphi^\veps) & = A_{ijkl}^\veps e_{kl}(\ub^\veps) - g_{kij}^\veps E_k(\vphi^\veps)\;,\\
        D_k^\veps(\ub^\veps,\vphi^\veps) & =  g_{kij}^\veps e_{ij}(\ub^\veps) + d_{kl}^\veps E_l(\vphi^\veps)\;,
    \end{split}
\end{equation}
which express the dependencies of the Cauchy stress tensor $\sigmab^\veps$ and
the electric displacement $\vec{D}^\veps$ on the strain tensor $\eb(\ub^\veps)
= {1\over2}\left(\nabla\ub^\veps + (\nabla\ub^\veps)^T\right)$, where
$\ub^\veps$ is the displacement field, and on the electric field
$\vec{E}(\vphi^\veps) = \nabla \vphi^\veps$, where $\vphi^\veps$ is the
electric potential. On the right hand side of (\ref{eq-constitutive}), we have
the fourth-order elastic tensor $A^\veps_{ijkl}$ ($A^\veps_{ijkl} =
A^\veps_{klij} = A^\veps_{jilk}$), the third-order tensor $g^\veps_{kij}$
($g^\veps_{kij} = g^\veps_{kji}$), which couples mechanical and electric
quantities, and the permeability tensor $d^\veps_{kl}$.

The quasi-static problem of the piezoelectric medium is given by the following
equilibrium equations
\begin{equation}\label{eq-equilibrium}
    \begin{split}
        -\nabla\cdot \sigmab^\veps(\ub^\veps,\vphi^\veps) & = \fb^\veps\;, \quad \mbox{ in }  \Omega_{mc}^\veps\;, \\
        -\nabla\cdot \vec D^\veps(\ub^\veps,\vphi^\veps) & = q_E^\veps\;,\quad \mbox{ in }  \Omega_m^\veps\;,
     \end{split}
\end{equation}
and by the boundary conditions
\begin{equation}\label{eq-bc}
    \begin{split}
        % \nb \cdot \sigmab^\veps & = \hb^\veps\;, \quad \mbox{ on } \Gamma_\sigma^\veps\;, \\
        % \nb \cdot \vec D^\veps & = \vrho_E^\veps\;, \quad \mbox{ on } \Gamma_{\vec D}^\veps\;, \\
        % \ub^\veps & = \bar\ub\;, \quad \mbox{ on } \Gamma_{\ub}^\veps\;, \\
        % \vphi^\veps & = \bar\vphi^0\;, \quad \mbox{ on } \Gamma_{\vphi}^\veps\;, \\
        \nb \cdot \sigmab^\veps = \hb^\veps\ \mbox{ on } \Gamma_\sigma^\veps\;,& \qquad
        \nb \cdot \vec D^\veps = \vrho_E^\veps\ \mbox{ on } \Gamma_{\vec D}^\veps\;, \\
        \ub^\veps = \bar\ub\ \mbox{ on } \Gamma_{\ub}^\veps\;,& \qquad
        \vphi^\veps = \bar\vphi\ \mbox{ on } \Gamma_{\vphi}^\veps\;, \\
    \end{split}
\end{equation}
where $\fb^\veps$, $\hb^\veps$ are the volume and surface forces, $q_E^\veps$,
$\vrho_E^\veps$ are the volume and surface charges and $\bar\ub$, $\bar\vphi$
are the prescribed displacements and electric potential, respectively. The
piezo-elastic medium occupies an open bounded region $\Omega \in \mathbb{R}^3$
which is decomposed into several non-overlapping parts: the piezoelectric
elastic matrix $\Omega_m^\veps$, conductive elastic parts
$\Omega_c^\veps = \bigcup_k \Omega_c^{k,\veps}$ and isolated void inclusions
$\Omega_o^\veps$, see~Fig.~\ref{fig-mac_mic}. The elastic part, i.e. the matrix
and the conductors, is denoted by
$\Omega_{mc}^\veps = \Omega_m^\veps \cup \Omega_c^\veps$.

The weak formulation of the problem stated above can be written as: Given
volume and surface forces $\fb^\veps$, $\hb^\veps$ and volume and surface
charges $q_E^\veps$, $\vrho_E^\veps$, find $\ub^\veps \in \Uspace(\Omega^\veps_{mc})$,
$\vphi^\veps \in {\mathcal V}(\Omega^\veps_{m})$ such
that for all $\vb \in \Uspace_0(\Omega^\veps_{mc})$, $\psi \in
{\mathcal V}_0(\Omega^\veps_{m})$
\begin{equation}\label{eq-waek_form}
    \begin{split}
        \int_{\Omega_m^\veps} & [\Ab^\veps \eeb{\ub^\veps} - (\gb^\veps)^T\cdot\nabla \vphi^\veps]: \eeb{\vb} \, \dV  + \int_{\Omega_c^\veps} [\Ab^\veps \eeb{\ub^\veps}]: \eeb{\vb}\,\dV  \\
         & = \int_{\Gamma_\sigma^\veps}\hb \cdot \vb\,\dS  + \int_{\Omega_{mc}^\veps} \fb^\veps \cdot \vb \,\dV \;, \\
        \int_{\Omega_m^\veps} & [\gb^\veps:\eeb{\ub^\veps} + \db^\veps \cdot\nabla \vphi^\veps]\cdot\nabla\psi \, \dV = \int_{\Gamma_{\vec D}^\veps}\vrho_E^\veps \psi\,\dS + \int_{\Omega_m^\veps}q_E^\veps\psi\,\dV \;.\\
    \end{split}
\end{equation}
Symbols $\Uspace$, $\Uspace_0$, ${\mathcal V}$, ${\mathcal
V}_0$ denote admissibility sets, where $\Uspace_0$, ${\mathcal V}_0$
are the sets with zero trace on the Dirichlet boundary. Further details can
be found in \cite{poropiezo2018}.

\subsection{Two-scale homogenization}

We apply the standard homogenization techniques, cf. \cite{Allaire1992} or
\cite{Cioranescu1999book}, to the problem (\ref{eq-waek_form}). It results in the limit
model for $\veps \longrightarrow 0$, where $\veps$ is the scale parameter
relating the microscopic and macroscopic length scales. The homogenization
process leads to local microscopic problems, defined within a
reference periodic cell, and to the global problem describing the behavior of
the homogenized medium at the macroscopic level. The global problem involves
the homogenized material coefficients which are evaluated using the solutions
of the local problems. Due to linearity of the problem, the microscopic and
macroscopic problems are decoupled.

As we assume given potentials $\bar\vphi^k$ in each of the electrode networks,
the dielectric properties must be appropriately rescaled in order to preserve
the finite electric field for the limit $\veps \longrightarrow 0$:
$\gb^\veps = \veps \bar\gb$, $\db^\veps = \veps^2 \bar\db$, cf.
\cite{poropiezo2018}.

\paragraph{Local problems and homogenized coeffficients}

The local microscopic responses of the piezoelectric structure are given by the
following sub-problems which are solved within the periodic reference cell $Y$,
see Fig.~\ref{fig-mac_mic}, that is decomposed similarly to the decomposition of
domain $\Omega$:
\begin{itemize}
    \item Find $\omegab^{ij}\in \Hspace(Y_{mc})$, $\eta^{ij}\in H_{\#0}^1(Y_m)$ such that for all
     $\vb \in \Hspace(Y_{mc})$, $\psi \in H_{\#0}^1(Y_m)$ and for any $i, j = 1, 2, 3$
    \begin{equation}\label{eq-mic1}
        \begin{split}
            \int_{Y_{mc}} \left[\Ab \eeb{\omegab^{ij} + \Pib^{ij}}\right]: \eeb{\vb}\,\dV  - \int_{Y_m} \left[\bar\gb^T\cdot\nabla \eta^{ij}\right]: \eeb{\vb} \, \dV &= 0\;, \\
            \int_{Y_m} \left[\bar\gb:\eeb{\omegab^{ij} + \Pib^{ij}} + \bar\db \cdot\nabla \eta^{ij}\right]\cdot\nabla\psi \, \dV &= 0\;,\\
        \end{split}
    \end{equation}
    where $\Pi^{ij}_k = y_j \delta_{ik}$.
    \item Find $\hat\omegab^k\in \Hspace(Y_{mc})$, $\hat\eta^k\in H_{\#0,k}^1(Y_m)$ such that for all
     $\vb \in \Hspace(Y_{mc})$, $\psi \in H_{\#0}^1(Y_m)$ and for any $k = 1, 2, \dots, k^c$ ($k^c$ is the number of conductors)
    \begin{equation}\label{eq-mic2}
        \begin{split}
            \int_{Y_{mc}} \left[\Ab \eeb{\hat\omegab^{k}}\right]: \eeb{\vb}\,\dV - \int_{Y_m} \left[\bar\gb^T\cdot\nabla \hat\eta^k\right]: \eeb{\vb} \,\dV &= 0\;, \\
            \int_{Y_m} \left[\bar\gb:\eeb{\hat\omegab^k} + \bar\db \cdot\nabla \hat\eta^k\right]\cdot\nabla\psi \,\dV &= 0\;.\\
        \end{split}
    \end{equation}
\end{itemize}
The microscopic sub-problems are solved with the periodic boundary conditions
and $\hat\eta^k = \delta_{ki}$ on $\Gamma^i_{mc}$ for $i = 1, 2, \dots, k^c$,
$\Gamma^i_{mc} = \overline{Y_m} \cap \overline{Y_c^i}$ is the interface between
the matrix part $Y_m$ and $i$-th conductor $Y_c^i$. By $\Hspace$ we refer to
the Sobolev space of $Y$-periodic functions, $H_{\#0,k}^1$ reflects the above
mentioned interface condition on $\Gamma^i_{mc}$ and $H_{\#0}^1$ is the set
of functions which are equal to zero on $\Gamma_{mc}$.

With the characteristic responses $\omegab^{ij}$, $\eta^{ij}$ and
$\hat\omegab^{k}$, $\hat\eta^{k}$ obtained by solving (\ref{eq-mic1}) and
(\ref{eq-mic2}), the homogenized material coefficients
$\Ab^H$ and $\Pb^{H,k}$ can be evaluated using the following expressions: % A, P
\begin{equation}\label{eq-coefs}
    \begin{split}
        A_{ijkl}^H & = {1\over \vert Y\vert} \left[\int_{Y_{mc}} \left[\Ab \eeb{\omegab^{ij} + \Pib^{ij}}\right]: \eeb{\omegab^{kl} + \Pib^{kl}}\,\dV
          + \int_{Y_m} \bar\db \nabla\eta^{ij} \cdot\nabla\eta^{kl}\,\dV\right]\;,\\
        %   - \int_{Y_m} \left[\bar\gb:\eeb{\Pib^{kl}}\right]\cdot\nabla\eta^{ij}\,\dV\right]\;,\\
        P^{H,k}_{ij} & = {1\over \vert Y\vert} \left[\int_{Y_{mc}} \left[\Ab \eeb{\hat\omegab^k}\right]: \eeb{\Pib^{ij}}\,\dV
          - \int_{Y_m} \left[\bar\gb:\eeb{\Pib^{ij}}\right]\cdot\nabla\hat\eta^k\,\dV\right]\;.\\
    \end{split}
\end{equation}

\paragraph{Macroscopic problem}

The global macroscopic problem is defined in terms of the homogenized
coefficients as: Find the macroscopic displacements $\ub^0 \in \Uspace
(\Omega)$ such that for all $\vb^0 \in \Uspace_0(\Omega)$
\begin{equation}\label{eq-mac}
    \int_{\Omega} [\Ab^H \eeb{\ub^0}]: \eeb{\vb}\,\dV
        = - \int_{\Omega} \eeb{\vb} : \sum\limits_k \Pb^{H,k} \bar\vphi^k \,\dV\;.
\end{equation}
We assumed that $\vrho_E = 0$, otherwise we would need an extra
coefficient related to the surface charge, see \cite{poropiezo2018}.

\subsection{Numerical simulation}

\begin{figure}
    \centering
    \includegraphics[width=0.7\linewidth]{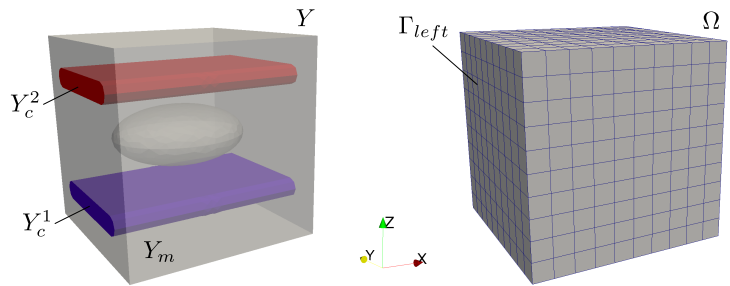}
    \caption{Left: the geometry of the reference periodic cell $Y$;
             Right: the macroscopic domain $\Omega$.}\label{fig-ex_mic_mac}
\end{figure}

In this section we illustrate the use of \SfePy{}'s homogenization engine in
the following setting. The macroscopic problem described by (\ref{eq-mac}) is
solved in the domain $\Omega$, depicted in Fig.~\ref{fig-ex_mic_mac} right,
that is fixed at its left face ($u^0_i = 0$ for $i = 1,2,3$ on
$\Gamma_{left}$). No volume and surface forces or charges are applied and the
deformation of the macroscopic sample is invoked only by the prescribed
electrical potential $\bar\vphi = \pm 10^4$\,V in the two embedded conductor
networks. The geometry of the representative volume element, which is used to
solve the microscopic problems (\ref{eq-mic1}), (\ref{eq-mic2}), is depicted in
Fig.~\ref{fig-ex_mic_mac} left. The material parameters of the piezoelectric
elastic matrix, made of barium-titanite, and metallic conductors are summarized
in Table~\ref{tab-matprop}.

\begin{table}[ht!]
    \begin{tabular}{lrrrrrr}
    \hline
    \vbox to 1.1em{}{\bf Piezoelectric matrix} & & & & & & \\
    Elasticity -- transverse isotropy [GPa]: & $A_{1111}$ & $A_{3333}$ & $A_{1122}$ & $A_{2233}$ & $A_{1313}$ & $A_{1212}$ \\
    ($A_{1111}=A_{2222}$, $A_{2233}=A_{1133}$, & 15.040 & 14.550 & 6.560 & 6.590 & 4.240 & 4.390 \\
    \ $A_{1313} = A_{2323}$) & & & & & &\\
    Piezo-coupling  [C/m$^2$]: & $g_{311}$ & $g_{322}$ & $g_{333}$ & $g_{223}$ & & \\
    ($g_{311}= g_{322}$, $g_{223}= g_{113}$) & -4.322 & -4.322 & 17.360 & 11.404 & & \\
    Dielectricity [$10^{-8}$ C/Vm]: & $d_{11}$ & $d_{33}$ & & & & \\
    ($d_{11} = d_{22}$)    & 1.284 & 1.505 & & & &\\
    \hline
    \vbox to 1.1em{}{\bf Metallic conductors} & & & & & & \\
    % Elasticity -- linear isotropy (GPa): & $A_{1111}$ & $A_{1122}$ & $A_{1212}$ & & & \\
    %       & 240.0 & 80.0 & 80.0 & & & \\
    Elasticity -- linear isotropy: & E [GPa] & $\nu$ [-] & & & & \\
    & 200.0 & 0.25 & & & & \\
    \hline
    \end{tabular}

    \caption{Properties of the piezoelectric matrix and metallic conductors.}\label{tab-matprop}
\end{table}

\begin{figure}
    \centering
    \includegraphics[width=0.44\linewidth]{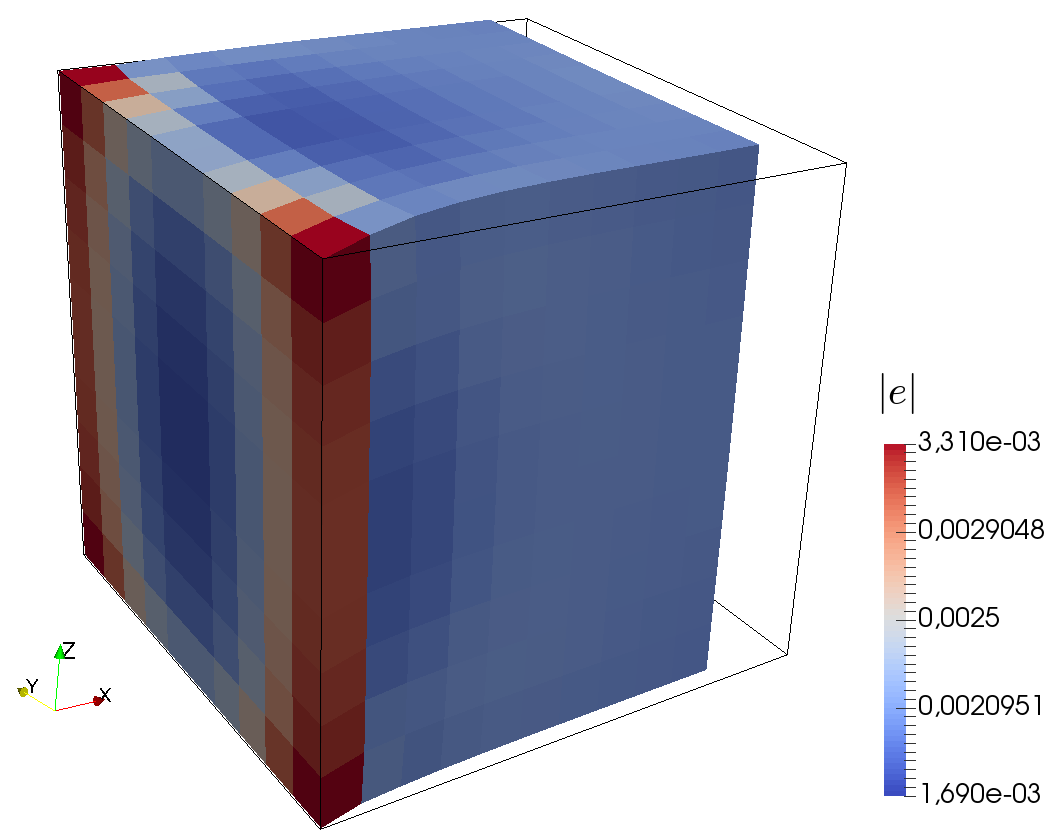}
    \caption{The deformed macroscopic sample (deformation scaled by factor 100)
      and the magnitude of macroscopic strain field.}\label{fig-ex_results1}
\end{figure}

\begin{figure}
    \centering
    \includegraphics[width=0.44\linewidth]{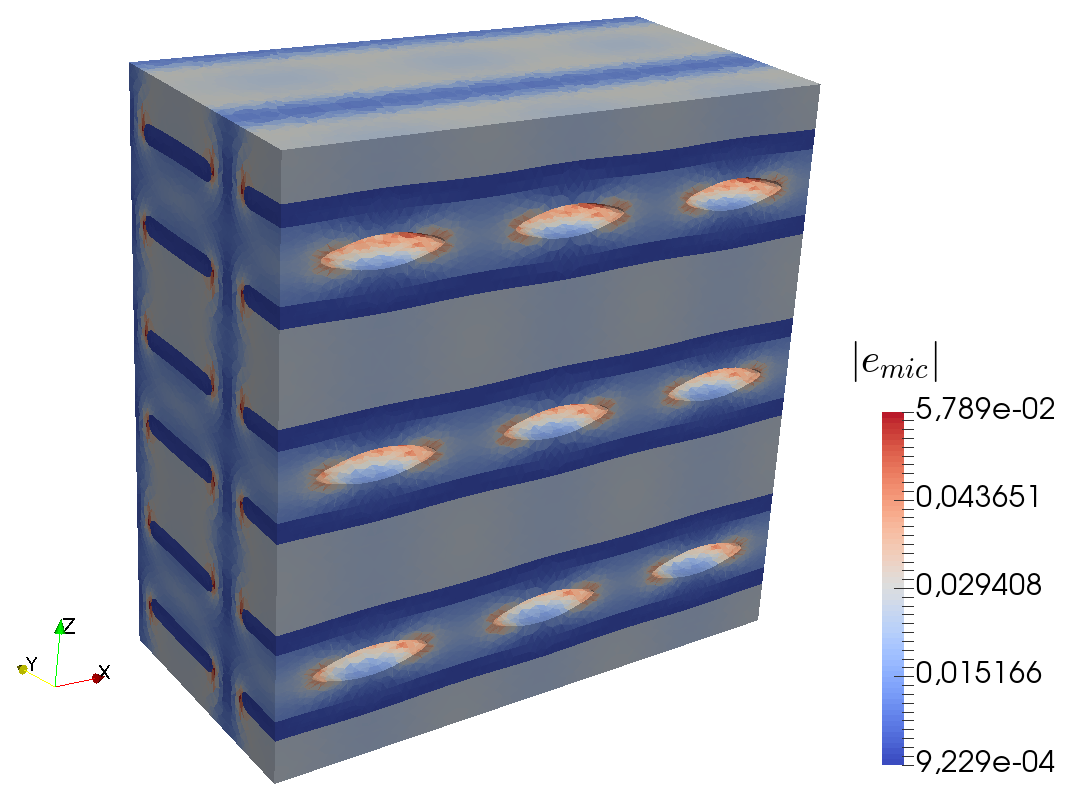}\hfill
    \includegraphics[width=0.44\linewidth]{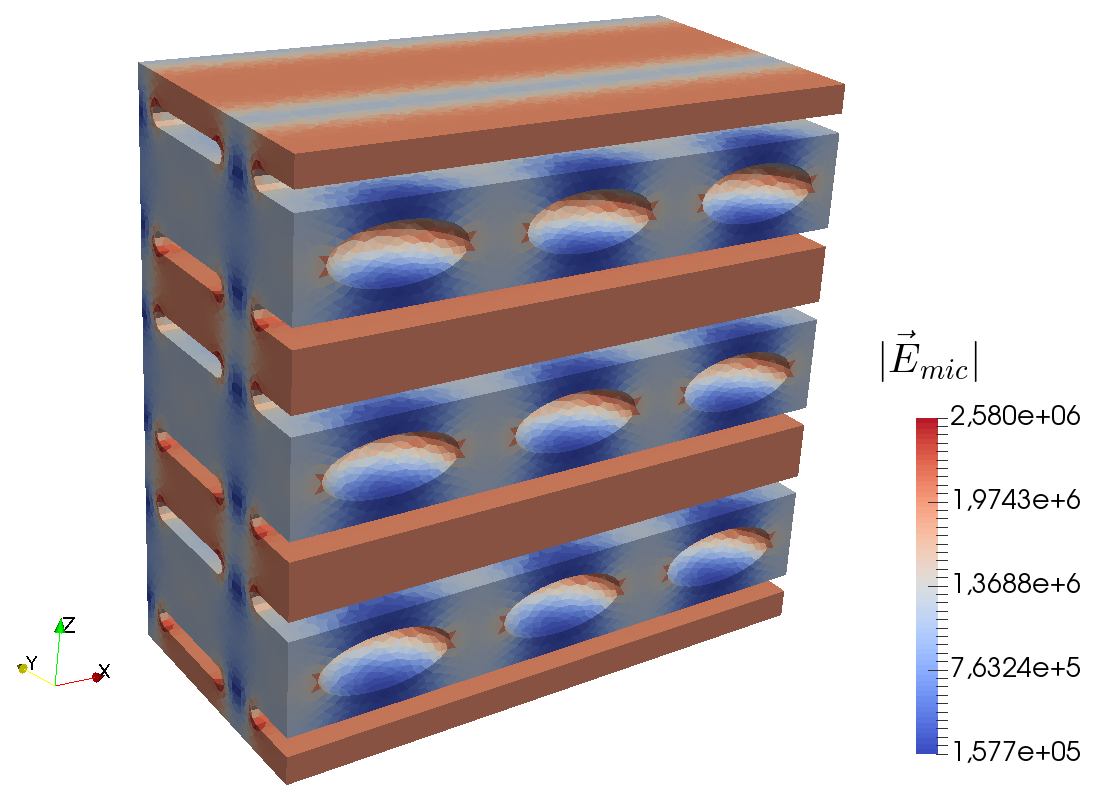}\\
    \caption{Left: the deformed microscopic structure (deformation scaled by
      factor 10) and the magnitude of reconstructed strain field; Right:
      the magnitude of the reconstructed electric field.}\label{fig-ex_results2}
\end{figure}

The results of the multiscale numerical simulation are shown in
Figs.~\ref{fig-ex_results1},~\ref{fig-ex_results2}. The macroscopic strain
field and the deformed macroscopic sample (deformation scaled by factor 100)
are presented in Fig.~\ref{fig-ex_results1}. Using the macroscopic solution
and the characteristic responses, we can reconstruct the fields at the
microscopic level for a given finite size $\veps_0$ in a chosen part of the
macroscopic domain. The reconstructed strain field
and the deformed microstructure (deformation scaled by factor 10) are shown in
Fig.~\ref{fig-ex_results2} left, the reconstructed electric field is depicted in
Fig.~\ref{fig-ex_results2} right. See \cite{poropiezo2018} for comparison of accuracy of
the reconstructed solutions with a full, much more demanding, simulation.
In the above article, the full (reference) simulation has approximately
$4.5 \times 10^5$ degrees of freedom while the microscopic problem
has only 741 DOFs and the macroscopic problem 577 DOFs.

\subsection{Multiscale analysis in \SfePy}

The linear multiscale analysis defined above is performed in \SfePy{} in two
steps:
\begin{enumerate}
    \item The local microscopic sub-problems (\ref{eq-mic1}), (\ref{eq-mic2})
    are solved using the homogenization engine, see Section~\ref{sec:hom_engine}. The
    engine is also used to evaluate the homogenized coefficients according to
    (\ref{eq-coefs}).
    \item The global macroscopic problem (\ref{eq-mac}) is solved, the known
      homogenized coefficients are employed.
\end{enumerate}

The definition of the global problem can be done in the similiar way as in the
simple heat conduction example presented in Section
\ref{sec:description_example}. The macroscopic equation in the declarative API
attains the form
\begin{Verbatim}[commandchars=\\\{\}]
\PY{n}{equations} \PY{o}{=} \PY{p}{\PYZob{}}
    \PY{l+s+s1}{\PYZsq{}}\PY{l+s+s1}{balance\PYZus{}of\PYZus{}forces}\PY{l+s+s1}{\PYZsq{}}\PY{p}{:} \PY{l+s+s2}{\PYZdq{}\PYZdq{}\PYZdq{}}\PY{l+s+s2}{dw\PYZus{}lin\PYZus{}elastic.i2.Omega(hom.A, v, u)}
\PY{l+s+s2}{                        = \PYZhy{} dw\PYZus{}lin\PYZus{}prestress.i2.Omega(hom.Pf, v)}\PY{l+s+s2}{\PYZdq{}\PYZdq{}\PYZdq{}}\PY{p}{,}
\PY{p}{\PYZcb{}}
\end{Verbatim}

where \texttt{hom.A} stands for the homogenized coefficients $\Ab^H$ and \texttt{hom.Pf}
is equal to $\sum_k \Pb^{H,k} \bar\vphi^k$, i.e. the sum of the
coefficients $\Pb^{H, k}$ multiplied by the prescribed electrical potentials $\bar\vphi^k$.
The homogenized material \texttt{hom} is declared as a function, which calls the
homogenization engine (via \SfePy{}'s built-in function \texttt{get\_homog\_coefs\_linear()}
in the following code) and returns the calculated homogenized parameters.
In the case of a linear problem, the same values are valid in all quadrature
points of a macroscopic domain (\texttt{coors} argument in the function below).
\begin{Verbatim}[commandchars=\\\{\}]
\PY{k}{def} \PY{n+nf}{get\PYZus{}homog\PYZus{}fun}\PY{p}{(}\PY{n}{ts}\PY{p}{,} \PY{n}{coors}\PY{p}{,} \PY{n}{mode}\PY{p}{,} \PY{o}{*}\PY{o}{*}\PY{n}{kwargs}\PY{p}{)}\PY{p}{:}
    \PY{o}{.}\PY{o}{.}\PY{o}{.}
    \PY{n}{coefs} \PY{o}{=} \PY{n}{get\PYZus{}homog\PYZus{}coefs\PYZus{}linear}\PY{p}{(}\PY{l+m+mi}{0}\PY{p}{,} \PY{l+m+mi}{0}\PY{p}{,} \PY{n+nb+bp}{None}\PY{p}{,}
                                   \PY{n}{micro\PYZus{}filename}\PY{o}{=}\PY{l+s+s1}{\PYZsq{}}\PY{l+s+s1}{piezo\PYZus{}elasticity\PYZus{}micro.py}\PY{l+s+s1}{\PYZsq{}}\PY{p}{,}
                                   \PY{n}{coefs\PYZus{}filename}\PY{o}{=}\PY{l+s+s1}{\PYZsq{}}\PY{l+s+s1}{coefs\PYZus{}piezo.h5}\PY{l+s+s1}{\PYZsq{}}\PY{p}{)}
    \PY{n}{Pf} \PY{o}{=} \PY{n}{coefs}\PY{p}{[}\PY{l+s+s1}{\PYZsq{}}\PY{l+s+s1}{P1}\PY{l+s+s1}{\PYZsq{}}\PY{p}{]} \PY{o}{*} \PY{n}{bar\PYZus{}phi}\PY{p}{[}\PY{l+m+mi}{0}\PY{p}{]} \PY{o}{+} \PY{n}{coefs}\PY{p}{[}\PY{l+s+s1}{\PYZsq{}}\PY{l+s+s1}{P2}\PY{l+s+s1}{\PYZsq{}}\PY{p}{]} \PY{o}{*} \PY{n}{bar\PYZus{}phi}\PY{p}{[}\PY{l+m+mi}{1}\PY{p}{]}
    \PY{n}{nqp} \PY{o}{=} \PY{n}{coors}\PY{o}{.}\PY{n}{shape}\PY{p}{[}\PY{l+m+mi}{0}\PY{p}{]}
    \PY{n}{out} \PY{o}{=} \PY{p}{\PYZob{}}
        \PY{l+s+s1}{\PYZsq{}}\PY{l+s+s1}{A}\PY{l+s+s1}{\PYZsq{}}\PY{p}{:} \PY{n}{np}\PY{o}{.}\PY{n}{tile}\PY{p}{(}\PY{n}{coefs}\PY{p}{[}\PY{l+s+s1}{\PYZsq{}}\PY{l+s+s1}{A}\PY{l+s+s1}{\PYZsq{}}\PY{p}{]}\PY{p}{,} \PY{p}{(}\PY{n}{nqp}\PY{p}{,} \PY{l+m+mi}{1}\PY{p}{,} \PY{l+m+mi}{1}\PY{p}{)}\PY{p}{)}\PY{p}{,}
        \PY{l+s+s1}{\PYZsq{}}\PY{l+s+s1}{Pf}\PY{l+s+s1}{\PYZsq{}}\PY{p}{:} \PY{n}{np}\PY{o}{.}\PY{n}{tile}\PY{p}{(}\PY{n}{Pf}\PY{p}{[}\PY{p}{:}\PY{p}{,} \PY{n}{np}\PY{o}{.}\PY{n}{newaxis}\PY{p}{]}\PY{p}{,} \PY{p}{(}\PY{n}{nqp}\PY{p}{,} \PY{l+m+mi}{1}\PY{p}{,} \PY{l+m+mi}{1}\PY{p}{)}\PY{p}{)}\PY{p}{,}
    \PY{p}{\PYZcb{}}
    \PY{k}{return} \PY{n}{out}
\PY{n}{functions} \PY{o}{=} \PY{p}{\PYZob{}}
    \PY{l+s+s1}{\PYZsq{}}\PY{l+s+s1}{get\PYZus{}homog}\PY{l+s+s1}{\PYZsq{}}\PY{p}{:} \PY{p}{(}\PY{n}{get\PYZus{}homog\PYZus{}fun}\PY{p}{,}\PY{p}{)}\PY{p}{,}
\PY{p}{\PYZcb{}}
\PY{n}{materials} \PY{o}{=} \PY{p}{\PYZob{}}
    \PY{l+s+s1}{\PYZsq{}}\PY{l+s+s1}{hom}\PY{l+s+s1}{\PYZsq{}}\PY{p}{:} \PY{l+s+s1}{\PYZsq{}}\PY{l+s+s1}{get\PYZus{}homog}\PY{l+s+s1}{\PYZsq{}}\PY{p}{,}
\PY{p}{\PYZcb{}}
\end{Verbatim}

To define the microscopic sub-problems which are solved by the homogenization engine
the following fields and variables are needed:
\begin{Verbatim}[commandchars=\\\{\}]
\PY{n}{fields} \PY{o}{=} \PY{p}{\PYZob{}}
    \PY{l+s+s1}{\PYZsq{}}\PY{l+s+s1}{displacement}\PY{l+s+s1}{\PYZsq{}}\PY{p}{:} \PY{p}{(}\PY{l+s+s1}{\PYZsq{}}\PY{l+s+s1}{real}\PY{l+s+s1}{\PYZsq{}}\PY{p}{,} \PY{l+s+s1}{\PYZsq{}}\PY{l+s+s1}{vector}\PY{l+s+s1}{\PYZsq{}}\PY{p}{,} \PY{l+s+s1}{\PYZsq{}}\PY{l+s+s1}{Ymc}\PY{l+s+s1}{\PYZsq{}}\PY{p}{,} \PY{l+m+mi}{1}\PY{p}{)}\PY{p}{,}
    \PY{l+s+s1}{\PYZsq{}}\PY{l+s+s1}{potential}\PY{l+s+s1}{\PYZsq{}}\PY{p}{:} \PY{p}{(}\PY{l+s+s1}{\PYZsq{}}\PY{l+s+s1}{real}\PY{l+s+s1}{\PYZsq{}}\PY{p}{,} \PY{l+s+s1}{\PYZsq{}}\PY{l+s+s1}{scalar}\PY{l+s+s1}{\PYZsq{}}\PY{p}{,} \PY{l+s+s1}{\PYZsq{}}\PY{l+s+s1}{Ym}\PY{l+s+s1}{\PYZsq{}}\PY{p}{,} \PY{l+m+mi}{1}\PY{p}{)}\PY{p}{,}
\PY{p}{\PYZcb{}}
\PY{n}{variables} \PY{o}{=} \PY{p}{\PYZob{}}
    \PY{c+c1}{\PYZsh{} displacement}
    \PY{l+s+s1}{\PYZsq{}}\PY{l+s+s1}{u}\PY{l+s+s1}{\PYZsq{}}\PY{p}{:} \PY{p}{(}\PY{l+s+s1}{\PYZsq{}}\PY{l+s+s1}{unknown field}\PY{l+s+s1}{\PYZsq{}}\PY{p}{,} \PY{l+s+s1}{\PYZsq{}}\PY{l+s+s1}{displacement}\PY{l+s+s1}{\PYZsq{}}\PY{p}{)}\PY{p}{,}
    \PY{l+s+s1}{\PYZsq{}}\PY{l+s+s1}{v}\PY{l+s+s1}{\PYZsq{}}\PY{p}{:} \PY{p}{(}\PY{l+s+s1}{\PYZsq{}}\PY{l+s+s1}{test field}\PY{l+s+s1}{\PYZsq{}}\PY{p}{,} \PY{l+s+s1}{\PYZsq{}}\PY{l+s+s1}{displacement}\PY{l+s+s1}{\PYZsq{}}\PY{p}{,} \PY{l+s+s1}{\PYZsq{}}\PY{l+s+s1}{u}\PY{l+s+s1}{\PYZsq{}}\PY{p}{)}\PY{p}{,}
    \PY{l+s+s1}{\PYZsq{}}\PY{l+s+s1}{Pi\PYZus{}u}\PY{l+s+s1}{\PYZsq{}}\PY{p}{:} \PY{p}{(}\PY{l+s+s1}{\PYZsq{}}\PY{l+s+s1}{parameter field}\PY{l+s+s1}{\PYZsq{}}\PY{p}{,} \PY{l+s+s1}{\PYZsq{}}\PY{l+s+s1}{displacement}\PY{l+s+s1}{\PYZsq{}}\PY{p}{,} \PY{l+s+s1}{\PYZsq{}}\PY{l+s+s1}{u}\PY{l+s+s1}{\PYZsq{}}\PY{p}{)}\PY{p}{,}
    \PY{l+s+s1}{\PYZsq{}}\PY{l+s+s1}{U1}\PY{l+s+s1}{\PYZsq{}}\PY{p}{:} \PY{p}{(}\PY{l+s+s1}{\PYZsq{}}\PY{l+s+s1}{parameter field}\PY{l+s+s1}{\PYZsq{}}\PY{p}{,} \PY{l+s+s1}{\PYZsq{}}\PY{l+s+s1}{displacement}\PY{l+s+s1}{\PYZsq{}}\PY{p}{,} \PY{l+s+s1}{\PYZsq{}}\PY{l+s+s1}{(set\PYZhy{}to\PYZhy{}None)}\PY{l+s+s1}{\PYZsq{}}\PY{p}{)}\PY{p}{,}
    \PY{l+s+s1}{\PYZsq{}}\PY{l+s+s1}{U2}\PY{l+s+s1}{\PYZsq{}}\PY{p}{:} \PY{p}{(}\PY{l+s+s1}{\PYZsq{}}\PY{l+s+s1}{parameter field}\PY{l+s+s1}{\PYZsq{}}\PY{p}{,} \PY{l+s+s1}{\PYZsq{}}\PY{l+s+s1}{displacement}\PY{l+s+s1}{\PYZsq{}}\PY{p}{,} \PY{l+s+s1}{\PYZsq{}}\PY{l+s+s1}{(set\PYZhy{}to\PYZhy{}None)}\PY{l+s+s1}{\PYZsq{}}\PY{p}{)}\PY{p}{,}
    \PY{c+c1}{\PYZsh{} potential}
    \PY{l+s+s1}{\PYZsq{}}\PY{l+s+s1}{r}\PY{l+s+s1}{\PYZsq{}}\PY{p}{:} \PY{p}{(}\PY{l+s+s1}{\PYZsq{}}\PY{l+s+s1}{unknown field}\PY{l+s+s1}{\PYZsq{}}\PY{p}{,} \PY{l+s+s1}{\PYZsq{}}\PY{l+s+s1}{potential}\PY{l+s+s1}{\PYZsq{}}\PY{p}{)}\PY{p}{,}
    \PY{l+s+s1}{\PYZsq{}}\PY{l+s+s1}{s}\PY{l+s+s1}{\PYZsq{}}\PY{p}{:} \PY{p}{(}\PY{l+s+s1}{\PYZsq{}}\PY{l+s+s1}{test field}\PY{l+s+s1}{\PYZsq{}}\PY{p}{,} \PY{l+s+s1}{\PYZsq{}}\PY{l+s+s1}{potential}\PY{l+s+s1}{\PYZsq{}}\PY{p}{,} \PY{l+s+s1}{\PYZsq{}}\PY{l+s+s1}{r}\PY{l+s+s1}{\PYZsq{}}\PY{p}{)}\PY{p}{,}
    \PY{l+s+s1}{\PYZsq{}}\PY{l+s+s1}{Pi\PYZus{}r}\PY{l+s+s1}{\PYZsq{}}\PY{p}{:} \PY{p}{(}\PY{l+s+s1}{\PYZsq{}}\PY{l+s+s1}{parameter field}\PY{l+s+s1}{\PYZsq{}}\PY{p}{,} \PY{l+s+s1}{\PYZsq{}}\PY{l+s+s1}{potential}\PY{l+s+s1}{\PYZsq{}}\PY{p}{,} \PY{l+s+s1}{\PYZsq{}}\PY{l+s+s1}{r}\PY{l+s+s1}{\PYZsq{}}\PY{p}{)}\PY{p}{,}
    \PY{l+s+s1}{\PYZsq{}}\PY{l+s+s1}{R1}\PY{l+s+s1}{\PYZsq{}}\PY{p}{:} \PY{p}{(}\PY{l+s+s1}{\PYZsq{}}\PY{l+s+s1}{parameter field}\PY{l+s+s1}{\PYZsq{}}\PY{p}{,} \PY{l+s+s1}{\PYZsq{}}\PY{l+s+s1}{potential}\PY{l+s+s1}{\PYZsq{}}\PY{p}{,} \PY{l+s+s1}{\PYZsq{}}\PY{l+s+s1}{(set\PYZhy{}to\PYZhy{}None)}\PY{l+s+s1}{\PYZsq{}}\PY{p}{)}\PY{p}{,}
    \PY{l+s+s1}{\PYZsq{}}\PY{l+s+s1}{R2}\PY{l+s+s1}{\PYZsq{}}\PY{p}{:} \PY{p}{(}\PY{l+s+s1}{\PYZsq{}}\PY{l+s+s1}{parameter field}\PY{l+s+s1}{\PYZsq{}}\PY{p}{,} \PY{l+s+s1}{\PYZsq{}}\PY{l+s+s1}{potential}\PY{l+s+s1}{\PYZsq{}}\PY{p}{,} \PY{l+s+s1}{\PYZsq{}}\PY{l+s+s1}{(set\PYZhy{}to\PYZhy{}None)}\PY{l+s+s1}{\PYZsq{}}\PY{p}{)}\PY{p}{,}
\PY{p}{\PYZcb{}}
\end{Verbatim}

The material parameters of the elastic matrix ($Y_m$) and the metalic conductors ($Y_c$)
are defined as follows, see Table~\ref{tab-matprop}:
\begin{Verbatim}[commandchars=\\\{\}]
\PY{n}{materials} \PY{o}{=} \PY{p}{\PYZob{}}
    \PY{l+s+s1}{\PYZsq{}}\PY{l+s+s1}{elastic}\PY{l+s+s1}{\PYZsq{}}\PY{p}{:} \PY{p}{(}\PY{p}{\PYZob{}}
        \PY{l+s+s1}{\PYZsq{}}\PY{l+s+s1}{D}\PY{l+s+s1}{\PYZsq{}}\PY{p}{:} \PY{p}{\PYZob{}}
            \PY{l+s+s1}{\PYZsq{}}\PY{l+s+s1}{Ym}\PY{l+s+s1}{\PYZsq{}}\PY{p}{:} \PY{n}{np}\PY{o}{.}\PY{n}{array}\PY{p}{(}\PY{p}{[}\PY{p}{[}\PY{l+m+mf}{1.504}\PY{p}{,} \PY{l+m+mf}{0.656}\PY{p}{,} \PY{l+m+mf}{0.659}\PY{p}{,} \PY{l+m+mi}{0}\PY{p}{,} \PY{l+m+mi}{0}\PY{p}{,} \PY{l+m+mi}{0}\PY{p}{]}\PY{p}{,}
                            \PY{p}{[}\PY{l+m+mf}{0.656}\PY{p}{,} \PY{l+m+mf}{1.504}\PY{p}{,} \PY{l+m+mf}{0.659}\PY{p}{,} \PY{l+m+mi}{0}\PY{p}{,} \PY{l+m+mi}{0}\PY{p}{,} \PY{l+m+mi}{0}\PY{p}{]}\PY{p}{,}
                            \PY{p}{[}\PY{l+m+mf}{0.659}\PY{p}{,} \PY{l+m+mf}{0.659}\PY{p}{,} \PY{l+m+mf}{1.455}\PY{p}{,} \PY{l+m+mi}{0}\PY{p}{,} \PY{l+m+mi}{0}\PY{p}{,} \PY{l+m+mi}{0}\PY{p}{]}\PY{p}{,}
                            \PY{p}{[}\PY{l+m+mi}{0}\PY{p}{,} \PY{l+m+mi}{0}\PY{p}{,} \PY{l+m+mi}{0}\PY{p}{,} \PY{l+m+mf}{0.424}\PY{p}{,} \PY{l+m+mi}{0}\PY{p}{,} \PY{l+m+mi}{0}\PY{p}{]}\PY{p}{,}
                            \PY{p}{[}\PY{l+m+mi}{0}\PY{p}{,} \PY{l+m+mi}{0}\PY{p}{,} \PY{l+m+mi}{0}\PY{p}{,} \PY{l+m+mi}{0}\PY{p}{,} \PY{l+m+mf}{0.439}\PY{p}{,} \PY{l+m+mi}{0}\PY{p}{]}\PY{p}{,}
                            \PY{p}{[}\PY{l+m+mi}{0}\PY{p}{,} \PY{l+m+mi}{0}\PY{p}{,} \PY{l+m+mi}{0}\PY{p}{,} \PY{l+m+mi}{0}\PY{p}{,} \PY{l+m+mi}{0}\PY{p}{,} \PY{l+m+mf}{0.439}\PY{p}{]}\PY{p}{]}\PY{p}{)} \PY{o}{*} \PY{l+m+mf}{1e11}\PY{p}{,}
            \PY{l+s+s1}{\PYZsq{}}\PY{l+s+s1}{Yc}\PY{l+s+s1}{\PYZsq{}}\PY{p}{:} \PY{n}{stiffness\PYZus{}from\PYZus{}youngpoisson}\PY{p}{(}\PY{l+m+mi}{3}\PY{p}{,} \PY{l+m+mf}{200e9}\PY{p}{,} \PY{l+m+mf}{0.25}\PY{p}{)}\PY{p}{\PYZcb{}}\PY{p}{\PYZcb{}}\PY{p}{,}\PY{p}{)}\PY{p}{,}
    \PY{l+s+s1}{\PYZsq{}}\PY{l+s+s1}{piezo}\PY{l+s+s1}{\PYZsq{}}\PY{p}{:} \PY{p}{(}\PY{p}{\PYZob{}}
        \PY{l+s+s1}{\PYZsq{}}\PY{l+s+s1}{g}\PY{l+s+s1}{\PYZsq{}}\PY{p}{:} \PY{n}{np}\PY{o}{.}\PY{n}{array}\PY{p}{(}\PY{p}{[}\PY{p}{[}\PY{l+m+mi}{0}\PY{p}{,} \PY{l+m+mi}{0}\PY{p}{,} \PY{l+m+mi}{0}\PY{p}{,} \PY{l+m+mi}{0}\PY{p}{,} \PY{l+m+mf}{11.404}\PY{p}{,} \PY{l+m+mi}{0}\PY{p}{]}\PY{p}{,}
                        \PY{p}{[}\PY{l+m+mi}{0}\PY{p}{,} \PY{l+m+mi}{0}\PY{p}{,} \PY{l+m+mi}{0}\PY{p}{,} \PY{l+m+mi}{0}\PY{p}{,} \PY{l+m+mi}{0}\PY{p}{,} \PY{l+m+mf}{11.404}\PY{p}{]}\PY{p}{,}
                        \PY{p}{[}\PY{o}{\PYZhy{}}\PY{l+m+mf}{4.322}\PY{p}{,} \PY{o}{\PYZhy{}}\PY{l+m+mf}{4.322}\PY{p}{,} \PY{l+m+mf}{17.360}\PY{p}{,} \PY{l+m+mi}{0}\PY{p}{,} \PY{l+m+mi}{0}\PY{p}{,} \PY{l+m+mi}{0}\PY{p}{]}\PY{p}{]}\PY{p}{)}\PY{p}{,}
        \PY{l+s+s1}{\PYZsq{}}\PY{l+s+s1}{d}\PY{l+s+s1}{\PYZsq{}}\PY{p}{:} \PY{n}{np}\PY{o}{.}\PY{n}{array}\PY{p}{(}\PY{p}{[}\PY{p}{[}\PY{l+m+mf}{1.284}\PY{p}{,} \PY{l+m+mi}{0}\PY{p}{,} \PY{l+m+mi}{0}\PY{p}{]}\PY{p}{,}
                        \PY{p}{[}\PY{l+m+mi}{0}\PY{p}{,} \PY{l+m+mf}{1.284}\PY{p}{,} \PY{l+m+mi}{0}\PY{p}{]}\PY{p}{,}
                        \PY{p}{[}\PY{l+m+mi}{0}\PY{p}{,} \PY{l+m+mi}{0}\PY{p}{,} \PY{l+m+mf}{1.505}\PY{p}{]}\PY{p}{]}\PY{p}{)} \PY{o}{*} \PY{l+m+mf}{1e\PYZhy{}8}\PY{p}{\PYZcb{}}\PY{p}{,}\PY{p}{)}\PY{p}{,}
\PY{p}{\PYZcb{}}
\end{Verbatim}

%
% The definition of the homogenized coefficients $\Ab^H$ defined in (\ref{\ref{eq-coefs}$_1$)
% consist of two parts and also their definition
The homogenized coefficients $\Ab^H$ can be introduced as
\begin{Verbatim}[commandchars=\\\{\}]
\PY{k+kn}{import} \PY{n+nn}{sfepy.homogenization.coefs\PYZus{}base} \PY{k+kn}{as} \PY{n+nn}{cb}
\PY{n}{coefs} \PY{o}{=} \PY{p}{\PYZob{}}
    \PY{l+s+s1}{\PYZsq{}}\PY{l+s+s1}{A1}\PY{l+s+s1}{\PYZsq{}}\PY{p}{:} \PY{p}{\PYZob{}}
        \PY{l+s+s1}{\PYZsq{}}\PY{l+s+s1}{requires}\PY{l+s+s1}{\PYZsq{}}\PY{p}{:} \PY{p}{[}\PY{l+s+s1}{\PYZsq{}}\PY{l+s+s1}{pis\PYZus{}u}\PY{l+s+s1}{\PYZsq{}}\PY{p}{,} \PY{l+s+s1}{\PYZsq{}}\PY{l+s+s1}{omega\PYZus{}ij}\PY{l+s+s1}{\PYZsq{}}\PY{p}{]}\PY{p}{,}
        \PY{l+s+s1}{\PYZsq{}}\PY{l+s+s1}{expression}\PY{l+s+s1}{\PYZsq{}}\PY{p}{:} \PY{l+s+s1}{\PYZsq{}}\PY{l+s+s1}{dw\PYZus{}lin\PYZus{}elastic.i2.Ymc(elastic.D, U1, U2)}\PY{l+s+s1}{\PYZsq{}}\PY{p}{,}
        \PY{l+s+s1}{\PYZsq{}}\PY{l+s+s1}{set\PYZus{}variables}\PY{l+s+s1}{\PYZsq{}}\PY{p}{:} \PY{p}{[}\PY{p}{(}\PY{l+s+s1}{\PYZsq{}}\PY{l+s+s1}{U1}\PY{l+s+s1}{\PYZsq{}}\PY{p}{,} \PY{p}{(}\PY{l+s+s1}{\PYZsq{}}\PY{l+s+s1}{omega\PYZus{}ij}\PY{l+s+s1}{\PYZsq{}}\PY{p}{,} \PY{l+s+s1}{\PYZsq{}}\PY{l+s+s1}{pis\PYZus{}u}\PY{l+s+s1}{\PYZsq{}}\PY{p}{)}\PY{p}{,} \PY{l+s+s1}{\PYZsq{}}\PY{l+s+s1}{u}\PY{l+s+s1}{\PYZsq{}}\PY{p}{)}\PY{p}{,}
                          \PY{p}{(}\PY{l+s+s1}{\PYZsq{}}\PY{l+s+s1}{U2}\PY{l+s+s1}{\PYZsq{}}\PY{p}{,} \PY{p}{(}\PY{l+s+s1}{\PYZsq{}}\PY{l+s+s1}{omega\PYZus{}ij}\PY{l+s+s1}{\PYZsq{}}\PY{p}{,} \PY{l+s+s1}{\PYZsq{}}\PY{l+s+s1}{pis\PYZus{}u}\PY{l+s+s1}{\PYZsq{}}\PY{p}{)}\PY{p}{,} \PY{l+s+s1}{\PYZsq{}}\PY{l+s+s1}{u}\PY{l+s+s1}{\PYZsq{}}\PY{p}{)}\PY{p}{]}\PY{p}{,}
        \PY{l+s+s1}{\PYZsq{}}\PY{l+s+s1}{class}\PY{l+s+s1}{\PYZsq{}}\PY{p}{:} \PY{n}{cb}\PY{o}{.}\PY{n}{CoefSymSym}\PY{p}{,}
    \PY{p}{\PYZcb{}}\PY{p}{,}
    \PY{l+s+s1}{\PYZsq{}}\PY{l+s+s1}{A2}\PY{l+s+s1}{\PYZsq{}}\PY{p}{:} \PY{p}{\PYZob{}}
        \PY{l+s+s1}{\PYZsq{}}\PY{l+s+s1}{requires}\PY{l+s+s1}{\PYZsq{}}\PY{p}{:} \PY{p}{[}\PY{l+s+s1}{\PYZsq{}}\PY{l+s+s1}{omega\PYZus{}ij}\PY{l+s+s1}{\PYZsq{}}\PY{p}{]}\PY{p}{,}
        \PY{l+s+s1}{\PYZsq{}}\PY{l+s+s1}{expression}\PY{l+s+s1}{\PYZsq{}}\PY{p}{:} \PY{l+s+s1}{\PYZsq{}}\PY{l+s+s1}{dw\PYZus{}diffusion.i2.Ym(piezo.d, R1, R2)}\PY{l+s+s1}{\PYZsq{}}\PY{p}{,}
        \PY{l+s+s1}{\PYZsq{}}\PY{l+s+s1}{set\PYZus{}variables}\PY{l+s+s1}{\PYZsq{}}\PY{p}{:} \PY{p}{[}\PY{p}{(}\PY{l+s+s1}{\PYZsq{}}\PY{l+s+s1}{R1}\PY{l+s+s1}{\PYZsq{}}\PY{p}{,} \PY{l+s+s1}{\PYZsq{}}\PY{l+s+s1}{omega\PYZus{}ij}\PY{l+s+s1}{\PYZsq{}}\PY{p}{,} \PY{l+s+s1}{\PYZsq{}}\PY{l+s+s1}{r}\PY{l+s+s1}{\PYZsq{}}\PY{p}{)}\PY{p}{,}
                          \PY{p}{(}\PY{l+s+s1}{\PYZsq{}}\PY{l+s+s1}{R2}\PY{l+s+s1}{\PYZsq{}}\PY{p}{,} \PY{l+s+s1}{\PYZsq{}}\PY{l+s+s1}{omega\PYZus{}ij}\PY{l+s+s1}{\PYZsq{}}\PY{p}{,} \PY{l+s+s1}{\PYZsq{}}\PY{l+s+s1}{r}\PY{l+s+s1}{\PYZsq{}}\PY{p}{)}\PY{p}{]}\PY{p}{,}
        \PY{l+s+s1}{\PYZsq{}}\PY{l+s+s1}{class}\PY{l+s+s1}{\PYZsq{}}\PY{p}{:} \PY{n}{cb}\PY{o}{.}\PY{n}{CoefSymSym}\PY{p}{,}
    \PY{p}{\PYZcb{}}\PY{p}{,}
    \PY{l+s+s1}{\PYZsq{}}\PY{l+s+s1}{A}\PY{l+s+s1}{\PYZsq{}}\PY{p}{:} \PY{p}{\PYZob{}}
        \PY{l+s+s1}{\PYZsq{}}\PY{l+s+s1}{requires}\PY{l+s+s1}{\PYZsq{}}\PY{p}{:} \PY{p}{[}\PY{l+s+s1}{\PYZsq{}}\PY{l+s+s1}{c.A1}\PY{l+s+s1}{\PYZsq{}}\PY{p}{,} \PY{l+s+s1}{\PYZsq{}}\PY{l+s+s1}{c.A2}\PY{l+s+s1}{\PYZsq{}}\PY{p}{]}\PY{p}{,}
        \PY{l+s+s1}{\PYZsq{}}\PY{l+s+s1}{expression}\PY{l+s+s1}{\PYZsq{}}\PY{p}{:} \PY{l+s+s1}{\PYZsq{}}\PY{l+s+s1}{c.A1 + c.A2}\PY{l+s+s1}{\PYZsq{}}\PY{p}{,}
        \PY{l+s+s1}{\PYZsq{}}\PY{l+s+s1}{class}\PY{l+s+s1}{\PYZsq{}}\PY{p}{:} \PY{n}{cb}\PY{o}{.}\PY{n}{CoefEval}\PY{p}{,}
    \PY{p}{\PYZcb{}}\PY{p}{,}
\PY{p}{\PYZcb{}}
\end{Verbatim}

where we follow the expression (\ref{eq-coefs}$_1$) which consists of two integrals
over domains $Y_{mc}$ (matrix + conductors) and $Y_c$ (conductors).
The definition of each coefficient has these parts:
\texttt{requires} -- the names of correctors needed for evaluation,
\texttt{expression} -- the expression to be evaluated,
\texttt{class} --  the coefficient class; it determines the way of evaluation
and the resulting matrix/array shape.
In our case, the class of \texttt{A1} and \texttt{A2} is \texttt{CoefSymSym}: it means
that the resulting coefficients are the fourth-order tensors in the symetric storage,
e.g. $sym \times sym$ matrices, where $sym$ is the number of components in
a symmetric stress/strain vector. Class \texttt{CoefEval} is used to evaluate
a simple mathematical expression, in our example, the summation of \texttt{A1}, \texttt{A2}.
In \texttt{set\_variables} section we say how to substitute the correctors into the variables
employed in the expression. For example, the code \texttt{('U1', ('omega\_ij', 'pis\_u'), 'u')}
is interpretted as: $U1 = \omegab^K + \Pib^K$, where $\omegab^K$ is stored in
\texttt{omega\_ij['u']}, $\Pib^K$ in \texttt{pis\_u['u']} and $K$ is
the multi-index attaining $11, 22, 33, 12, 13, 23$ for a 3D problem because of
the used \texttt{CoefSymSym} class.
In a similar way, the coefficients $\Pb^{H,1}$ can be introduced as
\begin{Verbatim}[commandchars=\\\{\}]
\PY{n}{coefs}\PY{o}{.}\PY{n}{update}\PY{p}{(}\PY{p}{\PYZob{}}
    \PY{l+s+s1}{\PYZsq{}}\PY{l+s+s1}{P1\PYZus{}1}\PY{l+s+s1}{\PYZsq{}}\PY{p}{:} \PY{p}{\PYZob{}}
        \PY{l+s+s1}{\PYZsq{}}\PY{l+s+s1}{requires}\PY{l+s+s1}{\PYZsq{}}\PY{p}{:} \PY{p}{[}\PY{l+s+s1}{\PYZsq{}}\PY{l+s+s1}{pis\PYZus{}u}\PY{l+s+s1}{\PYZsq{}}\PY{p}{,} \PY{l+s+s1}{\PYZsq{}}\PY{l+s+s1}{omega\PYZus{}k1}\PY{l+s+s1}{\PYZsq{}}\PY{p}{]}\PY{p}{,}
        \PY{l+s+s1}{\PYZsq{}}\PY{l+s+s1}{expression}\PY{l+s+s1}{\PYZsq{}}\PY{p}{:} \PY{l+s+s1}{\PYZsq{}}\PY{l+s+s1}{dw\PYZus{}lin\PYZus{}elastic.i2.Ymc(elastic.D, U1, U2)}\PY{l+s+s1}{\PYZsq{}}\PY{p}{,}
        \PY{l+s+s1}{\PYZsq{}}\PY{l+s+s1}{set\PYZus{}variables}\PY{l+s+s1}{\PYZsq{}}\PY{p}{:} \PY{p}{[}\PY{p}{(}\PY{l+s+s1}{\PYZsq{}}\PY{l+s+s1}{U1}\PY{l+s+s1}{\PYZsq{}}\PY{p}{,} \PY{l+s+s1}{\PYZsq{}}\PY{l+s+s1}{omega\PYZus{}k1}\PY{l+s+s1}{\PYZsq{}}\PY{p}{,} \PY{l+s+s1}{\PYZsq{}}\PY{l+s+s1}{u}\PY{l+s+s1}{\PYZsq{}}\PY{p}{)}\PY{p}{,}
                          \PY{p}{(}\PY{l+s+s1}{\PYZsq{}}\PY{l+s+s1}{U2}\PY{l+s+s1}{\PYZsq{}}\PY{p}{,} \PY{l+s+s1}{\PYZsq{}}\PY{l+s+s1}{pis\PYZus{}u}\PY{l+s+s1}{\PYZsq{}}\PY{p}{,} \PY{l+s+s1}{\PYZsq{}}\PY{l+s+s1}{u}\PY{l+s+s1}{\PYZsq{}}\PY{p}{)}\PY{p}{]}\PY{p}{,}
        \PY{l+s+s1}{\PYZsq{}}\PY{l+s+s1}{class}\PY{l+s+s1}{\PYZsq{}}\PY{p}{:} \PY{n}{cb}\PY{o}{.}\PY{n}{CoefSym}\PY{p}{,}
    \PY{p}{\PYZcb{}}\PY{p}{,}
    \PY{l+s+s1}{\PYZsq{}}\PY{l+s+s1}{P1\PYZus{}2}\PY{l+s+s1}{\PYZsq{}}\PY{p}{:} \PY{p}{\PYZob{}}
        \PY{l+s+s1}{\PYZsq{}}\PY{l+s+s1}{requires}\PY{l+s+s1}{\PYZsq{}}\PY{p}{:} \PY{p}{[}\PY{l+s+s1}{\PYZsq{}}\PY{l+s+s1}{pis\PYZus{}u}\PY{l+s+s1}{\PYZsq{}}\PY{p}{,} \PY{l+s+s1}{\PYZsq{}}\PY{l+s+s1}{omega\PYZus{}k1}\PY{l+s+s1}{\PYZsq{}}\PY{p}{]}\PY{p}{,}
        \PY{l+s+s1}{\PYZsq{}}\PY{l+s+s1}{expression}\PY{l+s+s1}{\PYZsq{}}\PY{p}{:} \PY{l+s+s1}{\PYZsq{}}\PY{l+s+s1}{dw\PYZus{}piezo\PYZus{}coupling.i2.Ym(piezo.g, U1, R1)}\PY{l+s+s1}{\PYZsq{}}\PY{p}{,}
        \PY{l+s+s1}{\PYZsq{}}\PY{l+s+s1}{set\PYZus{}variables}\PY{l+s+s1}{\PYZsq{}}\PY{p}{:} \PY{p}{[}\PY{p}{(}\PY{l+s+s1}{\PYZsq{}}\PY{l+s+s1}{R1}\PY{l+s+s1}{\PYZsq{}}\PY{p}{,} \PY{l+s+s1}{\PYZsq{}}\PY{l+s+s1}{omega\PYZus{}k1}\PY{l+s+s1}{\PYZsq{}}\PY{p}{,} \PY{l+s+s1}{\PYZsq{}}\PY{l+s+s1}{r}\PY{l+s+s1}{\PYZsq{}}\PY{p}{)}\PY{p}{,}
                          \PY{p}{(}\PY{l+s+s1}{\PYZsq{}}\PY{l+s+s1}{U1}\PY{l+s+s1}{\PYZsq{}}\PY{p}{,} \PY{l+s+s1}{\PYZsq{}}\PY{l+s+s1}{pis\PYZus{}u}\PY{l+s+s1}{\PYZsq{}}\PY{p}{,} \PY{l+s+s1}{\PYZsq{}}\PY{l+s+s1}{u}\PY{l+s+s1}{\PYZsq{}}\PY{p}{)}\PY{p}{]}\PY{p}{,}
        \PY{l+s+s1}{\PYZsq{}}\PY{l+s+s1}{class}\PY{l+s+s1}{\PYZsq{}}\PY{p}{:} \PY{n}{cb}\PY{o}{.}\PY{n}{CoefSym}\PY{p}{,}
    \PY{p}{\PYZcb{}}\PY{p}{,}
    \PY{l+s+s1}{\PYZsq{}}\PY{l+s+s1}{P1}\PY{l+s+s1}{\PYZsq{}}\PY{p}{:} \PY{p}{\PYZob{}}
        \PY{l+s+s1}{\PYZsq{}}\PY{l+s+s1}{requires}\PY{l+s+s1}{\PYZsq{}}\PY{p}{:} \PY{p}{[}\PY{l+s+s1}{\PYZsq{}}\PY{l+s+s1}{c.P1\PYZus{}1}\PY{l+s+s1}{\PYZsq{}}\PY{p}{,} \PY{l+s+s1}{\PYZsq{}}\PY{l+s+s1}{c.P1\PYZus{}2}\PY{l+s+s1}{\PYZsq{}}\PY{p}{]}\PY{p}{,}
        \PY{l+s+s1}{\PYZsq{}}\PY{l+s+s1}{expression}\PY{l+s+s1}{\PYZsq{}}\PY{p}{:} \PY{l+s+s1}{\PYZsq{}}\PY{l+s+s1}{c.P1\PYZus{}1 \PYZhy{} c.P1\PYZus{}2}\PY{l+s+s1}{\PYZsq{}}\PY{p}{,}
        \PY{l+s+s1}{\PYZsq{}}\PY{l+s+s1}{class}\PY{l+s+s1}{\PYZsq{}}\PY{p}{:} \PY{n}{cb}\PY{o}{.}\PY{n}{CoefEval}\PY{p}{,}
    \PY{p}{\PYZcb{}}\PY{p}{,}
\PY{p}{\PYZcb{}}\PY{p}{)}
\end{Verbatim}

Here, the \texttt{CoefSym} class is employed due to the second-order coefficient
which can be represented as a vector with dimension $sym$.

The required correctors \texttt{omega\_ij}, see (\ref{eq-mic1}), and \texttt{pis\_u}
are defined as follows:
\begin{Verbatim}[commandchars=\\\{\}]
\PY{n}{requirements} \PY{o}{=} \PY{p}{\PYZob{}}
    \PY{l+s+s1}{\PYZsq{}}\PY{l+s+s1}{pis\PYZus{}u}\PY{l+s+s1}{\PYZsq{}}\PY{p}{:} \PY{p}{\PYZob{}}
        \PY{l+s+s1}{\PYZsq{}}\PY{l+s+s1}{variables}\PY{l+s+s1}{\PYZsq{}}\PY{p}{:} \PY{p}{[}\PY{l+s+s1}{\PYZsq{}}\PY{l+s+s1}{u}\PY{l+s+s1}{\PYZsq{}}\PY{p}{]}\PY{p}{,}
        \PY{l+s+s1}{\PYZsq{}}\PY{l+s+s1}{class}\PY{l+s+s1}{\PYZsq{}}\PY{p}{:} \PY{n}{cb}\PY{o}{.}\PY{n}{ShapeDimDim}\PY{p}{,}
    \PY{p}{\PYZcb{}}\PY{p}{,}
    \PY{l+s+s1}{\PYZsq{}}\PY{l+s+s1}{omega\PYZus{}ij}\PY{l+s+s1}{\PYZsq{}}\PY{p}{:} \PY{p}{\PYZob{}}
        \PY{l+s+s1}{\PYZsq{}}\PY{l+s+s1}{requires}\PY{l+s+s1}{\PYZsq{}}\PY{p}{:} \PY{p}{[}\PY{l+s+s1}{\PYZsq{}}\PY{l+s+s1}{pis\PYZus{}u}\PY{l+s+s1}{\PYZsq{}}\PY{p}{]}\PY{p}{,}
        \PY{l+s+s1}{\PYZsq{}}\PY{l+s+s1}{ebcs}\PY{l+s+s1}{\PYZsq{}}\PY{p}{:} \PY{p}{[}\PY{l+s+s1}{\PYZsq{}}\PY{l+s+s1}{fixed\PYZus{}u}\PY{l+s+s1}{\PYZsq{}}\PY{p}{,} \PY{l+s+s1}{\PYZsq{}}\PY{l+s+s1}{fixed\PYZus{}r}\PY{l+s+s1}{\PYZsq{}}\PY{p}{]}\PY{p}{,}
        \PY{l+s+s1}{\PYZsq{}}\PY{l+s+s1}{epbcs}\PY{l+s+s1}{\PYZsq{}}\PY{p}{:} \PY{p}{[}\PY{l+s+s1}{\PYZsq{}}\PY{l+s+s1}{p\PYZus{}ux}\PY{l+s+s1}{\PYZsq{}}\PY{p}{,} \PY{l+s+s1}{\PYZsq{}}\PY{l+s+s1}{p\PYZus{}uy}\PY{l+s+s1}{\PYZsq{}}\PY{p}{,} \PY{l+s+s1}{\PYZsq{}}\PY{l+s+s1}{p\PYZus{}uz}\PY{l+s+s1}{\PYZsq{}}\PY{p}{,} \PY{l+s+s1}{\PYZsq{}}\PY{l+s+s1}{p\PYZus{}rx}\PY{l+s+s1}{\PYZsq{}}\PY{p}{,} \PY{l+s+s1}{\PYZsq{}}\PY{l+s+s1}{p\PYZus{}ry}\PY{l+s+s1}{\PYZsq{}}\PY{p}{,} \PY{l+s+s1}{\PYZsq{}}\PY{l+s+s1}{p\PYZus{}rz}\PY{l+s+s1}{\PYZsq{}}\PY{p}{]}\PY{p}{,}
        \PY{l+s+s1}{\PYZsq{}}\PY{l+s+s1}{equations}\PY{l+s+s1}{\PYZsq{}}\PY{p}{:} \PY{p}{\PYZob{}}
            \PY{l+s+s1}{\PYZsq{}}\PY{l+s+s1}{eq1}\PY{l+s+s1}{\PYZsq{}}\PY{p}{:} \PY{l+s+s2}{\PYZdq{}\PYZdq{}\PYZdq{}}\PY{l+s+s2}{dw\PYZus{}lin\PYZus{}elastic.i2.Ymc(elastic.D, v, u)}
\PY{l+s+s2}{                    \PYZhy{} dw\PYZus{}piezo\PYZus{}coupling.i2.Ym(piezo.g, v, r)}
\PY{l+s+s2}{                   = \PYZhy{}dw\PYZus{}lin\PYZus{}elastic.i2.Ymc(elastic.D, v, Pi\PYZus{}u)}\PY{l+s+s2}{\PYZdq{}\PYZdq{}\PYZdq{}}\PY{p}{,}
            \PY{l+s+s1}{\PYZsq{}}\PY{l+s+s1}{eq2}\PY{l+s+s1}{\PYZsq{}}\PY{p}{:} \PY{l+s+s2}{\PYZdq{}\PYZdq{}\PYZdq{}}\PY{l+s+s2}{dw\PYZus{}piezo\PYZus{}coupling.i2.Ym(piezo.g, u, s)}
\PY{l+s+s2}{                    + dw\PYZus{}diffusion.i2.Ym(piezo.d, s, r)}
\PY{l+s+s2}{                   = \PYZhy{}dw\PYZus{}piezo\PYZus{}coupling.i2.Ym(piezo.g, Pi\PYZus{}u, s)}\PY{l+s+s2}{\PYZdq{}\PYZdq{}\PYZdq{}}\PY{p}{,}
        \PY{p}{\PYZcb{}}\PY{p}{,}
        \PY{l+s+s1}{\PYZsq{}}\PY{l+s+s1}{set\PYZus{}variables}\PY{l+s+s1}{\PYZsq{}}\PY{p}{:} \PY{p}{[}\PY{p}{(}\PY{l+s+s1}{\PYZsq{}}\PY{l+s+s1}{Pi\PYZus{}u}\PY{l+s+s1}{\PYZsq{}}\PY{p}{,} \PY{l+s+s1}{\PYZsq{}}\PY{l+s+s1}{pis\PYZus{}u}\PY{l+s+s1}{\PYZsq{}}\PY{p}{,} \PY{l+s+s1}{\PYZsq{}}\PY{l+s+s1}{u}\PY{l+s+s1}{\PYZsq{}}\PY{p}{)}\PY{p}{]}\PY{p}{,}
        \PY{l+s+s1}{\PYZsq{}}\PY{l+s+s1}{class}\PY{l+s+s1}{\PYZsq{}}\PY{p}{:} \PY{n}{cb}\PY{o}{.}\PY{n}{CorrDimDim}\PY{p}{,}
        \PY{l+s+s1}{\PYZsq{}}\PY{l+s+s1}{dump\PYZus{}variables}\PY{l+s+s1}{\PYZsq{}}\PY{p}{:} \PY{p}{[}\PY{l+s+s1}{\PYZsq{}}\PY{l+s+s1}{u}\PY{l+s+s1}{\PYZsq{}}\PY{p}{,} \PY{l+s+s1}{\PYZsq{}}\PY{l+s+s1}{r}\PY{l+s+s1}{\PYZsq{}}\PY{p}{]}\PY{p}{,}
    \PY{p}{\PYZcb{}}\PY{p}{,}
\PY{p}{\PYZcb{}}
\end{Verbatim}

The class \texttt{ShapeDimDim} is used to define the symbol $\Pi^{ij}_k = y_j \delta_{ik}$
and \texttt{CorrDimDim} ensures the corrector with $dim \times dim$ components,
$dim$ is the space dimension. Note that a corrector can also depend on
another corrector as in the code above, where \texttt{pis\_u} is required to solve
\texttt{omega\_ij}. The correctors introduced in (\ref{eq-mic2}) can be defined
in \SfePy{} as
\begin{Verbatim}[commandchars=\\\{\}]
\PY{n}{requirements}\PY{o}{.}\PY{n}{update}\PY{p}{(}\PY{p}{\PYZob{}}
    \PY{l+s+s1}{\PYZsq{}}\PY{l+s+s1}{omega\PYZus{}k1: \PYZob{}}
        \PY{l+s+s1}{\PYZsq{}}\PY{l+s+s1}{requires}\PY{l+s+s1}{\PYZsq{}}\PY{p}{:} \PY{p}{[}\PY{l+s+s1}{\PYZsq{}}\PY{l+s+s1}{pis\PYZus{}r}\PY{l+s+s1}{\PYZsq{}}\PY{p}{]}\PY{p}{,}
        \PY{l+s+s1}{\PYZsq{}}\PY{l+s+s1}{ebcs}\PY{l+s+s1}{\PYZsq{}}\PY{p}{:} \PY{p}{[}\PY{l+s+s1}{\PYZsq{}}\PY{l+s+s1}{fixed\PYZus{}u}\PY{l+s+s1}{\PYZsq{}}\PY{p}{,} \PY{l+s+s1}{\PYZsq{}}\PY{l+s+s1}{fixed\PYZus{}interface}\PY{l+s+s1}{\PYZsq{}}\PY{p}{]}\PY{p}{,}
        \PY{l+s+s1}{\PYZsq{}}\PY{l+s+s1}{epbcs}\PY{l+s+s1}{\PYZsq{}}\PY{p}{:} \PY{p}{[}\PY{l+s+s1}{\PYZsq{}}\PY{l+s+s1}{p\PYZus{}ux}\PY{l+s+s1}{\PYZsq{}}\PY{p}{,} \PY{l+s+s1}{\PYZsq{}}\PY{l+s+s1}{p\PYZus{}uy}\PY{l+s+s1}{\PYZsq{}}\PY{p}{,} \PY{l+s+s1}{\PYZsq{}}\PY{l+s+s1}{p\PYZus{}uz}\PY{l+s+s1}{\PYZsq{}}\PY{p}{,} \PY{l+s+s1}{\PYZsq{}}\PY{l+s+s1}{p\PYZus{}rx}\PY{l+s+s1}{\PYZsq{}}\PY{p}{,} \PY{l+s+s1}{\PYZsq{}}\PY{l+s+s1}{p\PYZus{}ry}\PY{l+s+s1}{\PYZsq{}}\PY{p}{,} \PY{l+s+s1}{\PYZsq{}}\PY{l+s+s1}{p\PYZus{}rz}\PY{l+s+s1}{\PYZsq{}}\PY{p}{]}\PY{p}{,}
        \PY{l+s+s1}{\PYZsq{}}\PY{l+s+s1}{equations}\PY{l+s+s1}{\PYZsq{}}\PY{p}{:} \PY{p}{\PYZob{}}
            \PY{l+s+s1}{\PYZsq{}}\PY{l+s+s1}{eq1}\PY{l+s+s1}{\PYZsq{}}\PY{p}{:} \PY{l+s+s2}{\PYZdq{}\PYZdq{}\PYZdq{}}\PY{l+s+s2}{dw\PYZus{}lin\PYZus{}elastic.i2.Ymc(elastic.D, v, u)}
\PY{l+s+s2}{                    \PYZhy{} dw\PYZus{}piezo\PYZus{}coupling.i2.Ym(piezo.g, v, r) = 0}\PY{l+s+s2}{\PYZdq{}\PYZdq{}\PYZdq{}}\PY{p}{,}
            \PY{l+s+s1}{\PYZsq{}}\PY{l+s+s1}{eq2}\PY{l+s+s1}{\PYZsq{}}\PY{p}{:} \PY{l+s+s2}{\PYZdq{}\PYZdq{}\PYZdq{}}\PY{l+s+s2}{dw\PYZus{}piezo\PYZus{}coupling.i2.Ym(piezo.g, u, s)}
\PY{l+s+s2}{                    + dw\PYZus{}diffusion.i2.Ym(piezo.d, s, r) = 0}\PY{l+s+s2}{\PYZdq{}\PYZdq{}\PYZdq{}}
        \PY{p}{\PYZcb{}}\PY{p}{,}
        \PY{l+s+s1}{\PYZsq{}}\PY{l+s+s1}{class}\PY{l+s+s1}{\PYZsq{}}\PY{p}{:} \PY{n}{cb}\PY{o}{.}\PY{n}{CorrOne}\PY{p}{,}
        \PY{l+s+s1}{\PYZsq{}}\PY{l+s+s1}{dump\PYZus{}variables}\PY{l+s+s1}{\PYZsq{}}\PY{p}{:} \PY{p}{[}\PY{l+s+s1}{\PYZsq{}}\PY{l+s+s1}{u}\PY{l+s+s1}{\PYZsq{}}\PY{p}{,} \PY{l+s+s1}{\PYZsq{}}\PY{l+s+s1}{r}\PY{l+s+s1}{\PYZsq{}}\PY{p}{]}\PY{p}{,}
    \PY{p}{\PYZcb{}}\PY{p}{,}
\PY{p}{\PYZcb{}}\PY{p}{)}
\end{Verbatim}

The class \texttt{CoefOne} corresponds to the scalar corrector function.
The correctors are solved with the periodic boundary conditions defined in the
lines with the keyword \texttt{epbcs} and the Dirichlet (essential) boundary
conditions defined in the lines with the keyword \texttt{ebcs}.

The multiscale simulation can be run by calling the \texttt{simple.py} script,
see Section~\ref{sec:description_run}, with the name of the description file
for the macroscopic problem as a script parameter. The script runs the
simulation at the macroscopic level and invokes the homogenization engine
through the material function. The full sources of this example can be found in
the \SfePy{} package in \texttt{examples/multiphysics/}:
\texttt{piezo\_elasticity\_macro.py} defines the macroscopic problem, and
\texttt{piezo\_elasticity\_micro.py} defines the computations on the reference
periodic cell of the microstructure. For the version of the sources used in this
article see \cite{sfepy-examples-zenodo}.

\section{Conclusion}
\label{conclusion}

We introduced the open source finite element package \SfePy{}, a code written
(mostly) in Python for solving various kinds of problems described by partial
differential equations and discretized by the finite element method. The design
of the code was discussed and illustrated using a simple heat conduction
example.

Special attention was devoted to the description of the \SfePy{}'s
homogenization engine, a sub-package for defining complex multiscale problems.
This feature was introduced in a tutorial-like form using a multiscale numerical
simulation of a piezoelectric structure.

For the complete code of the examples presented, together with the required FE
meshes and the 2018.3 version of \SfePy{}, see \cite{sfepy-examples-zenodo}.
Further documentation and many more examples of \SfePy{} use can be found on
the project's web site \cite{sfepy-web}.

\paragraph{Acknowledgment}
% \begin{acknowledgements}
This work was supported by the projects GA17-12925S and GA16-03823S
of the Czech Science Foundation and by the project LO1506 of the Czech Ministry
of Education, Youth and Sports.
% \end{acknowledgements}

\bibliographystyle{spbasic}
\bibliography{sfepy-bibliography}

\end{document}